# Heat kernel methods in finance: the SABR model

Carmelo Vaccaro

The work presented in this report has been carried out with the support of Reuters Financial Software (Puteaux, France) and under the direction of Adrien Bourgerie, to whom the author is particularly grateful.

For remarks, comments or propositions of collaboration, please write me at carmelo.vaccaro@sfr.fr

**Abstract**

The SABR model is a stochastic volatility model not admitting a closed form solution. Hagan, Kumar, Leniewski and Woodward [HaKuLeWo] have obtained an approximate solution by means of perturbative techniques. A more precise approximation was found by Henry-Labordère ([H-L 2005] and [H-L 2008]) with the *heat kernel expansion* method. The latter relies on deep and hard theorems from Riemannian geometry which are almost totally unknown to the professionals of finance, who however are those primarily interested in these results.

The goal of this report is to fill this gap and to make these topics understandable with a basic knowledge of calculus and linear algebra.

# Table of contents

# 1 Presentation of the report

The *Black-Scholes-Merton (BSM) model*, introduced in [Black-Scholes] and [Merton], is one of the most used financial models for pricing options. However it has been the target of many critics, some even considering it to be the cause of the October 1987 market crash [Bouchaud].

The main critic to the BSM model is that it is a hyper-simplified and unrealistically simple model: indeed the BSM model assumes that the volatility driving an asset is constant and that the price of the latter follows a log-normal process, facts that are contradicted by real market data.

Another critic is that the BSM implied volatility, which is obtained on the assumption that the volatility be constant, is indeed a non-constant function of strike and time to maturity. This engenders a contradiction that makes the use of BSM implied volatilities conceptually flawed.

Moreover, since any implied volatility gives rise to a different dynamic of the asset, when using the BSM model one is assuming that the same underlying asset follows different dynamics, one for each (vanilla) option on it, and this fact is very hard to justify.

Finally since it is the option on the underlying asset that determines the implied volatility of the latter, in the BSM framework it is then the option that affects the underlying, while in reality it is exactly the opposite that is true.

The BSM model is also criticized on practical grounds. Suppose for instance that we are using the BSM model for pricing a barrier option; since the latter has two strikes (one for the option in itself and one for the barrier), we wonder which one of the two BSM implied volatilities must be used, or if it must be used a combination of the two.

A model that circumvents the latter inconsistencies is the *local volatility model*, introduced in 1994 in [Dupire] and [Derman-Kani]. This model assumes that the volatility of the underlying asset is no longer constant but depends on time and on the value of the underlying. The local volatility model presents however an inconsistency with real market data: as observed in [HaKuLeWo], it predicts a dynamic behavior of smiles and skews which is exactly opposite to that observed in the market. Indeed when the price of the underlying decreases, the local volatility model predicts that the smile shifts to higher prices; when the price increases, the model predicts that the smile shifts to lower prices. In reality, asset prices and market smiles move in the same direction. This contradiction between the model and the marketplace tends to destabilize the Delta and Vega hedges, which often perform worse than the BSM hedges.

A class of models that avoids all these problems is that of *stochastic volatility models*, that is models of diffusion in which the volatility is assumed to be stochastic. Examples of these models are the *Hull-White two-factor model*, the *Heston model* and the *SABR model*. In this report we will concentrate on the latter.

The SABR model has been introduced by Hagan, Kumar, Leniewski and Woodward in 2002 [HaKuLeWo]. We will not discuss in details the reasons why it is a good model (a thorough discussion of this subject can be found in the introduction of [ReMcWhi]); we only say here that the SABR model gives quite consistent hedges with a manageable computational cost, that is it is an excellent compromise between precision and easiness to use.

To obtain a solution to the option pricing problem in the SABR model, the key fact is that with any system of $n$ stochastic differential equations we can associate a *differentiable manifold* $M$ of dimension $n$ such that the Kolmogoroff backward equation verified by its probability density is a heat equation on $M$. In particular for the SABR model, $n=2$ and $M$ is the Poincaré plane. Since the latter is a very studied mathematical object, the solution to the option pricing problem can be achieved with no much difficulty.

We observe that unlike the Heston model, the above solution is not exact: in [HaKuLeWo] and [HaLeWo], the authors have found an approximate solution by using perturbative methods (the *Varadhan's theorem*).

A more powerful technique is that of the *heat kernel expansion*: the solution to the heat equation is expressed as a power series whose coefficients can be computed using geometric information on $M$. Henry-Labordère has used the latter at the first order ([H-L 2005] and [H-L 2008]) to obtain a more accurate solution than that of [HaKuLeWo] and [HaLeWo]; Paulot has used the expansion at the second order [Paulot] to obtain an even more precise solution.

In fact the heat kernel expansion method can be applied to a general financial model involving a system of stochastic differential equations. It represents a new technique for solving pricing equations in finance, which until now, except the case where a closed form solution existed, could only be studied with computationally costly procedures. The heat kernel expansion method gives in any case a *semi-closed form solution* (i.e., a solution written as a power series) which can be obtained with geometric techniques. The stochastic differential problem translates then to a geometric problem for the manifold $M$.

Two references for these techniques are the books [Avramidi] and [H-L 2008] (where besides the SABR are also presented applications to multi-asset options and to the LMM model).

## 1.1 THE LINK WITH RIEMANNIAN GEOMETRY AND HEAT KERNEL EXPANSIONS

A *Riemannian structure* on $IR^n$ is a differentiable function that associates with any element of $IR^n$ a positive definite square matrix of order $n$. Since there is equivalence between scalar products and positive definite matrices, this means that a Riemannian structure is a way to define in a differentiable manner a metric that can change in each point.

For instance, let us associate the identity matrix with every point of $IR^n$. Since the identity matrix gives the usual Euclidean metric, in this case we have the standard metric that we use in calculus. The goal of Riemannian geometry is to consider a general metric and define accordingly the operators of derivative, differential, integral, etc…

One of the main differences between the general case and the Euclidean one is that the matrix of the metric is not necessarily constant. For instance, we will see that the metric for the SABR model is the *Poincaré plane* defined in a point $(x, y)$ by the matrix

$$\begin{pmatrix} \frac{1}{y^2} & 0 \\ 0 & \frac{1}{y^2} \end{pmatrix}.$$

The SABR consists of a system of two correlated stochastic differential equations

$$\begin{cases} F_t(\omega) = F_0 + \int_0^t A_u(\omega) F_u^\beta(\omega) dW_u(\omega) \\ A_t(\omega) = \alpha + \nu \int_0^t A_u(\omega) dZ_u(\omega) \\ corr(W_t, Z_t) = \rho \end{cases},$$

where $F_0, \beta, \alpha, \nu$ and $\rho$ are input constants. To show the link with Riemannian geometry and heat kernel expansions we consider a general system of $n$ correlated stochastic differential equations,

$$\begin{cases} X_i(t,\omega) = \alpha_i + \int_0^t b_i(u, X(u,\omega)) du + \int_0^t \sigma_i(u, X(u,\omega)) dW_u(\omega), & i = 1, \cdots, n \\ corr(X_i(t), X_j(t)) = \rho_{ij}(t), & i, j = 1, \cdots, n \end{cases}$$

where $X = (X_1, \cdots, X_n)$. Let us given a European option $C$ on a (multidimensional) asset whose dynamic is described by the preceding equations. Let $T$ be the maturity of $C$ and let the payoff at $T$ be a function $f(X_1(T), \cdots, X_n(T))$. The fair value of $C$ verifies the equation

$$\frac{\partial C}{\partial T}(\alpha, t, T) = \int_{-\infty}^{+\infty} f(x) p(T, x, t, \alpha) dx,$$

where $p$ is the probability density of the system, $x = (x_1, \cdots, x_n)$ and $\alpha = (\alpha_1, \cdots, \alpha_n)$. Let us set $\tau = T - t$ and $p(\tau, x, \alpha) := p(T, x, t, \alpha)$; then $p$ verifies the Kolmogoroff backward equation

$$\frac{\partial p}{\partial \tau} = Dp,$$

with

$$D = \sum_{i=1}^n b_i(\tau, \alpha) \frac{\partial}{\partial \alpha_i} + \sum_{i,j=1}^n \frac{1}{2} \rho_{ij}(\tau) \sigma_i(\tau, \alpha) \sigma_j(\tau, \alpha) \frac{\partial^2}{\partial \alpha_i \partial \alpha_j}$$

and with initial condition

$$p(0, x, \alpha) = \delta(x - \alpha)$$

($\delta$ is the Dirac function defined by $\delta(z) = 1$ for $z = 0$ and $\delta(z) = 0$ otherwise). This equation is a heat equation. Moreover we have that the matrix

$$g(x) = \left( \frac{1}{2} \rho_{ij}(\tau) \sigma_i(\tau, \alpha) \sigma_j(\tau, \alpha) \right)_{i,j=1,\cdots,n}$$

is positive definite. If the functions $\sigma_i$ and $\rho_{ij}$ are differentiable, then $g$ is differentiable and thus the above system of stochastic differential equations determines a Riemannian structure with respect to which the Kolmogoroff backward equation is a heat equation. In [HaKuLeWo] and [HaLeWo], Hagan, Kumar, Leniewski and Woodward have found an approximate solution by means of perturbative techniques. A more powerful method than the latter and giving a more accurate approximation is that of the *heat kernel expansion*, the presentation of which is the goal of this report.

## 1.2 CONTRIBUTIONS OF THIS REPORT

As already said, new powerful techniques are available for solving option pricing problems, techniques taken from advanced Riemannian geometry. People working in finance (and potentially interested in these results) have little or no knowledge of Riemannian geometry, which is a very complicated and hard mathematical field for which one has to spend a lot of time and effort to master only the most elementary facts.

This report aims at being a gentle introduction that could allow a non-specialist to read the more advanced papers and books [H-L 2005], [H-L 2008], [Paulot], [Avramidi]. The primary goal is that of giving a presentation of the topic that could be rigorous and complete and at the same time easy to understand and not too abstract. This would not have been possible if we had followed the usual approach to present this subject (for instance [Lee]).

Indeed Riemannian geometry has been employed until now only in mathematics and physics applications. Unlike these two fields, the Riemannian manifolds encountered in financial modeling are much simpler and reduce to the space $I\!R^n$. This manifold is trivial from the differential geometric point of view[1] and has little or no interest to a mathematician or physicist. So when one is narrowed to financial applications, the standard general way of presenting Riemannian manifolds is a useless complication.

In this report we have rewritten all the definitions and theorems for the special case of the manifold $I\!R^n$. This accomplishes our goal since it reduces the difficulty of the subject to that of elementary calculus and linear algebra. This work is an original one since there is no such an approach in any book or account on Riemannian geometry.

---

[1] its atlas reduces to a single chart

There has been in recent times a growing interest in Riemannian geometric techniques for solving financial problems ([Avramidi], [Farinelli], [H-L 2008]), so the utility of this report goes beyond the SABR model.

The report is organized as follows: Chapter 2 presents the basic notions of Riemannian geometry, Chapter 3 treats the Poincaré plane (which is the space model used for the SABR) and finally Chapter 4 applies all these notions for the case of the SABR.

# 2  Introduction to Riemannian geometry

## 2.1 REMINDERS FROM CALCULUS

Let $n$ be a natural number, let $M \subset I\!R^n$ and let $\phi : M \to R$ be a function. Let $\{e_1, \cdots, e_n\}$ be the canonical base of $I\!R^n$, that is for $i \in \{1, \cdots, n\}$, $e_i$ is the vector whose $i$-th component is 1 and the other are zero. Let $x \in M$; the $i$-*th partial derivative of* $\phi$ *at* $x$, denoted $\partial_i \phi(x)$ or $\partial_i \phi_x$, is the derivative at $x$ of $\phi$ when restricted to the straight line passing trough $x$ and having $e_i$ as direction, that is to $\{x + \alpha \, e_i : \alpha \in I\!R\}$.

Let $\phi$ admit $i$-th partial derivatives at $x$ for every $i \in \{1, \cdots, n\}$; the *gradient of* $\phi$ *at* $x$ is the vector function $(\partial_1 \phi(x), \cdots, \partial_n \phi(x))$ and is denoted $grad\,\phi(x)$.

Let $v = (v_1, \cdots, v_n)$ be a vector of $I\!R^n$ of norm 1; the *directional derivative of* $\phi$ *at* $x$ *along* $v$, denoted $D_v \phi(x)$ or $D_v \phi_x$, is the derivative at $x$ of $\phi$ when restricted to $\{x + \alpha v : \alpha \in I\!R\}$. It is well known from basic calculus that

$$D_v \phi(x) = v_1 \partial_1 \phi(x) + \cdots + v_n \partial_n \phi(x),$$

that is the directional derivative of $\phi$ along $v$ is equal to the scalar product between $v$ and *grad* $\phi$.

Let $i, j \in \{1, \cdots, n\}$ and suppose that the $j-$th derivative of $\phi$ admits $i$-th partial derivative. Then we define the $(i, j) -$ *second partial derivative of* $\phi$ as $\partial_i(\partial_j \phi)$ and we denote it $\partial_{ij} \phi$.

Let $\phi_1, \cdots, \phi_n : M \to I\!R$ be such that $\phi_i$ admits $i$-th partial derivative at $x$ and let $\Phi = (\phi_1, \cdots, \phi_n)$ be the vector function whose components are the $\phi_i$. The *divergence of* $\Phi$ *at* $x$ is defined as

$$\text{div } \Phi(x) = \sum_{i=1}^{n} \partial_i \phi_i(x).$$

Let $\phi : M \to I\!R$ admit second partial derivatives at $x$; the *Laplace operator* (or *Laplacian*) of $\phi$ at $x$ is defined as

$$\Delta \phi(x) := \sum_{i=1}^{n} \partial_{ii} \phi(x),$$

that is the Laplacian of $\phi$ is the divergence of the gradient of $\phi$.

Now let $\psi$, $b^1, \cdots, b^n$ and $g^{ij}$ for $i, j \in \{1, \cdots, n\}$ be functions from $M$ to $I\!R$; the operator sending a function $\phi : M \to I\!R$ admitting second partial derivatives to the function

$$\sum_{i,j=1}^{n} g^{ij} \partial_{ij} \phi + \sum_{h=1}^{n} b^h \partial_h \phi + \psi$$

is called a *second order linear differential operator*. This operator is said *elliptic* if for any non-zero $x = (x_1, \cdots, x_n) \in M$ we have

$$\sum_{i,j=1}^{n} x_i x_j g^{ij}(x) \neq 0.$$

In particular the Laplacian is elliptic.

## 2.2 PROPERTIES OF DIRECTIONAL DERIVATIVES

In this section we will extend the concept of directional derivatives to a more general context. First we take $v$ as a not necessarily constant vector but instead as a differentiable function $v : M \to I\!R^n$. In particular $v(x)$ is not necessarily of norm 1. Thus we define

$$D_v \phi(x) := v(x) \cdot grad\ \phi(x)\ ,$$

where $v(x) \cdot grad\ \phi(x)$ denotes the scalar product between these two vectors.

Let $Y$ be a function from $IR^n$ to $IR^m$ and let $Y_1, \cdots, Y_m$ be its scalar coordinates; then $D_v Y$ is the function $(D_v Y_1, \cdots, D_v Y_m)$, that is

$$D_v Y(x) = (v(x) \cdot grad\ Y_1(x), \cdots, v(x) \cdot grad\ Y_m(x))$$

and written as a linear combination in the canonical basis $\{e_1, \cdots, e_m\}$ of $IR^m$

$$D_v Y(x) = \sum_{k=1}^{m} (v(x) \cdot grad\ Y_k(x)) e_k\ .$$

Let us call *vector field on* $IR^n$ any differentiable function from $IR^n$ to $IR^n$ and let us denote $Vec(IR^n)$ the set of all vector fields. We will denote $Diff(IR^n, IR^m)$ the set of differentiable functions from $IR^n$ to $IR^m$. In particular

$$Vec(IR^n) = Diff(IR^n, IR^n)\ .$$

If $f: IR^n \to IR$ and $X: IR^n \to IR^m$ are functions then as usual $fX$ is the function sending $x \in IR^n$ to $f(x)X(x)$. If $f$ and $X$ are differentiable, then so is $fX$. In particular for any real $\alpha$, we denote $\alpha X$ the function sending $x \in IR^n$ to $\alpha X(x)$.

We prove that the directional derivative verifies the following properties:

1) for every $X_1, X_2 \in Vec(IR^n)$, $Y \in Diff(IR^n, IR^m)$ and $f_1, f_2 : IR^n \to IR$ differentiable functions we have

$$D_{f_1 X_1 + f_2 X_2} Y = f_1 D_{X_1} Y + f_2 D_{X_2} Y\ ;$$

2) for every $X \in Vec(IR^n)$, $Y_1, Y_2 \in Diff(IR^n, IR^m)$ and $\alpha_1, \alpha_2 \in IR$ we have

$$D_X (\alpha_1 Y_1 + \alpha_2 Y_2) = \alpha_1 D_X Y_1 + \alpha_2 D_X Y_2\ ;$$

3) for every $X \in Vec(IR^n)$, $Y \in Diff(IR^n, IR^m)$, for every $f: IR^n \to IR$ differentiable function and $x \in IR^n$ we have

$$D_X (fY)_x = f(x) D_X Y(x) + (X(x) \cdot grad\ f(x)) Y(x)\ .$$

The first two properties are obvious consequences of the linearity of scalar products of vectors and linearity of the gradient with respect to scalars. Let us prove the third. We have that

$$D_X(fY)_x = \left(D_X(fY_i)_x\right)_{i=1,\cdots,n}$$

and using the Leibniz rule that

$$D_X(fY_i)_x = X(x)\cdot grad(fY_i)_x = X(x)\cdot\left(f(x)\,grad\,Y_i(x) + Y_i(x)\,grad\,f(x)\right) =$$

$$= f(x)\left(X(x)\cdot grad\,Y_i(x)\right) + Y_i(x)\left(X(x)\cdot grad\,f(x)\right),$$

thus

$$D_X(fY)_x = \left(f(x)\left(X(x)\cdot grad\,Y_i(x)\right) + Y_i(x)\left(X(x)\cdot grad\,f(x)\right)\right)_{i=1,\cdots,m} =$$

$$= f(x)\left(X(x)\cdot grad\,Y_i(x)\right)_{i=1,\cdots,m} + \left(X(x)\cdot grad\,f(x)\right)\left(Y_i(x)\right)_{i=1,\cdots,m} =$$

$$= f(x)D_X Y(x) + \left(X(x)\cdot grad\,f(x)\right)Y(x).$$

## 2.3 CONNECTIONS

We now define an operator which generalizes derivatives by taking the three properties of the preceding section as axioms. Let $n$ and $m$ be natural numbers; a $(n,m)-connection$ is a function $\nabla : Vec(IR^n) \times Diff(IR^n, IR^m) \to Diff(IR^n, IR^m)$ such that, using the notation $\nabla_X Y$ instead of $\nabla(X,Y)$ we have:

1) for every $X_1, X_2 \in Vec(IR^n)$, $Y \in Diff(IR^n, IR^m)$ and $f_1, f_2 : IR^n \to IR$ differentiable functions we have

$$\nabla_{f_1 X_1 + f_2 X_2} Y = f_1 \nabla_{X_1} Y + f_2 \nabla_{X_2} Y ;$$

2) for every $X \in Vec(IR^n)$, $Y_1, Y_2 \in Diff(IR^n, IR^m)$ and $\alpha_1, \alpha_2 \in IR$ we have

$$\nabla_X(\alpha_1 Y_1 + \alpha_2 Y_2) = \alpha_1 \nabla_X Y_1 + \alpha_2 \nabla_X Y_2 ;$$

3) for every $X \in Vec(IR^n)$, $Y \in Diff(IR^n, IR^m)$, for every $f: IR^n \to IR$ differentiable function and $x \in IR^n$ we have

$$\nabla_X (fY)_x = f(x)\nabla_X Y(x) + (X(x) \cdot grad\, f(x))Y(x).$$

Let us fix a basis $\{e_1, \cdots, e_n\}$ of $IR^n$ and for $i = 1, \cdots, n$ let us denote $E_i$ the function sending any element of $IR^n$ to $e_i$. Fix also a basis $\{e'_1, \cdots, e'_m\}$ of $IR^m$ and for $j = 1, \cdots, m$ denote $E'_j$ the function sending any element of $IR^n$ to $e'_j$.

Let us given a connection $\nabla$ and $i \in \{1, \cdots, n\}$, $j \in \{1, \cdots, m\}$ we have that $\nabla_{E_i} E'_j(x)$ is a linear combination of elements of $\{e'_1, \cdots, e'_m\}$, so there exist real numbers $\Gamma_{ij}^k(x)$ depending on $x \in IR^n$ for $k = 1, \cdots, m$ such that

$$\nabla_{E_i} E'_j(x) = \sum_{k=1}^{m} \Gamma_{ij}^k(x) e'_k.$$

The coefficients $\Gamma_{ij}^k(x)$ are called *the Christoffel symbols of the connection* $\nabla$.

If $\nabla$ is the standard directional derivative, then the Christoffel symbols are all zero. Indeed since $E'_j$ is a constant function, its gradient is zero thus the directional derivative of $E'_j$ is zero and therefore its coordinates in any base are zero.

Let $X$ be a vector field. Thus for $x \in IR^n$, $X(x)$ belongs to $IR^n$ thus it is a linear combination of $e_1, \cdots, e_n$. Therefore there exist $X_1, \cdots, X_n$ functions from $IR^n$ to $IR$ such that $X(x) = \sum_{i=1}^{n} X_i(x) e_i$. Thus by recalling that $E_i$ is the vector field sending any element of $IR^n$ to $e_i$, we have that $X(x) = \sum_{i=1}^{n} X_i(x) E_i(x)$, that is

$$X = \sum_{i=1}^{n} X_i E_i.$$

In the same way, given a differentiable function $Y$ from $IR^n$ to $IR^m$ there exist $Y_1, \cdots, Y_m$ functions from $IR^n$ to $IR$ such that

$$Y = \sum_{j=1}^{m} Y_j E'_j.$$

We have the following result:

**Theorem.** *Let $\nabla$ be a connection on $IR^n$, let $X \in Vec(IR^n)$, $Y \in Diff(IR^n, IR^m)$ and let $X = \sum_{i=1}^{n} X_i E_i$, $Y = \sum_{j=1}^{m} Y_j E'_j$. Then*

$$\nabla_X Y(x) = \sum_{k=1}^{n} \left( X(x) \cdot grad\, Y_k(x) + \sum_{i=1}^{n} \sum_{j=1}^{m} X_i(x) Y_j(x) \Gamma_{ij}^{k}(x) \right) e'_k.$$

**Proof** By using properties 1 and 2 of connections, we have

$$\nabla_X Y(x) = \nabla_{\sum_{i=1}^{n} X_i E_i} \left( \sum_{j=1}^{m} Y_j E'_j \right)_x = \sum_{i=1}^{n} \sum_{j=1}^{m} X_i(x) \nabla_{E_i} \left( Y_j E'_j \right)_x.$$

By using property 3 of connections and the decomposition using Christoffel symbols, we have that

$$\nabla_{E_i} \left( Y_j E'_j \right)_x = Y_j(x) \nabla_{E_i} E'_j(x) + \left( E_i(x) \cdot grad\, Y_j(x) \right) E'_j(x) =$$
$$= Y_j(x) \sum_{k=1}^{m} \Gamma_{ij}^{k}(x) e'_k + \left( E_i(x) \cdot grad\, Y_j(x) \right) e'_j,$$

thus $\nabla_X Y(x)$ is equal to

$$\sum_{i=1}^{n} \sum_{j=1}^{m} X_i(x) \left[ Y_j(x) \sum_{k=1}^{m} \Gamma_{ij}^{k}(x) e'_k + \left( E_i(x) \cdot grad\, Y_j(x) \right) e'_j \right] =$$
$$= \sum_{i=1}^{n} \sum_{j,k=1}^{m} X_i(x) Y_j(x) \Gamma_{ij}^{k}(x) e'_k(x) + \sum_{i=1}^{n} \sum_{j=1}^{m} X_i(x) \left( e_i \cdot grad\, Y_j(x) \right) e'_j.$$

Finally we have that

$$\sum_{i=1}^{n} \sum_{j=1}^{m} X_i(x) \left( e_i \cdot grad\, Y_j(x) \right) e'_j = \sum_{j=1}^{m} \left[ \sum_{i=1}^{n} \left( X_i(x) e_i \cdot grad\, Y_j(x) \right) \right] e'_j =$$
$$= \sum_{j=1}^{m} \left[ \left( \sum_{i=1}^{n} X_i(x) e_i \right) \cdot grad\, Y_j(x) \right] e'_j = \sum_{j=1}^{m} \left( X(x) \cdot grad\, Y_j(x) \right) e'_j,$$

and replacing the index $j$ with the index $k$ we have the claim. ∎

**Remark** Let $\Gamma^k(x)$ be the matrix $\left(\Gamma_{ij}^k(x)\right)_{j=1,\cdots,m}^{i=1,\cdots,n}$. Thus the preceding equation can be written in the following form

$$\nabla_X Y(p) = \sum_{k=1}^{m} \left( X(x) \cdot grad\, Y_k(x) + X(x)\Gamma^k(x)Y(x) \right) e'_k.$$

**Remark** Let $X = E_i$ for $i \in \{1,\cdots,n\}$ and let us denote $\nabla_i := \nabla_{E_i}$. Then we have

$$\nabla_i Y(x) = \sum_{k=1}^{m} \left( \partial_i Y_k(x) + \sum_{j=1}^{m} \Gamma_{ij}^k(x) Y_j(x) \right) e'_k.$$

## 2.4 THE COVARIANT DERIVATIVE

A *(twice differentiable) curve* on $IR^n$ is a twice differentiable function $\phi: I \to IR^n$, where $I$ is an interval of real numbers. We denote $\phi_i$ for $i = 1,\cdots,n$ the components of $\phi$, that is for $t \in I$ we have that $\phi(t)$ is the vector $(\phi_1(t),\cdots,\phi_n(t))$. We denote $\dot\phi$ the first derivative of $\phi$ and $\ddot\phi$ the second.

Let $m$ be a natural number and let us denote $Diff_m(\phi)$ the set of all differentiable functions $V: I \to IR^m$. Linear combinations of elements of $Diff_m(\phi)$ are defined in the same way as for vector fields on $IR^n$.

Let $V \in Diff_m(\phi)$ and let $\overline{V} \in Diff(IR^n, IR^m)$; we say that $V$ is *extendible to* $\overline{V}$ if for every $t \in I$ we have $V(t) = \overline{V}(\phi(t))$.

Given a basis $\{e'_1, \cdots, e'_m\}$ of $IR^m$ we denote $E'_j$ for $j = 1, \cdots, m$ the element of $Diff_m(\phi)$ sending any $t \in I$ to $e'_j$. As for vector fields on $IR^n$ we have that if $V \in Diff_m(\phi)$ there exist functions $V_1, \cdots, V_m$ from $I$ to $IR$ such that

$$V = \sum_{j=1}^{m} V_j E'_j.$$

Let $\overline{E'}_j$ be the element of $Diff(IR^n, IR^m)$ sending any $x \in IR^n$ to $e'_j$; then $E'_j$ is extendible to $\overline{E'}_j$.

Suppose given a curve $\phi : I \to IR^n$ and a $(n,m)$-connection $\nabla$. Let $\{e_1, \cdots, e_n\}$ be a basis of $IR^n$ and let $E_i : IR^n \to IR^n$ for $i = 1, \cdots, n$ be the function sending any $x \in IR^n$ to $e_i$.

A *covariant derivative along $\phi$ with respect to* $\nabla$ is a function

$$Der_\phi : Diff(IR^n, IR^m) \to Diff(IR^n, IR^m)$$

verifying the following properties:

1) for every $V_1, V_2 \in Diff_m(\phi)$ and $\alpha_1, \alpha_2 \in IR$ we have

$$Der_\phi(\alpha_1 V_1 + \alpha_2 V_2) = \alpha_1 Der_\phi(V_1) + \alpha_2 Der_\phi(V_2);$$

2) for every $V \in Diff_m(\phi)$ and $f : IR^n \to IR$ differentiable function we have

$$Der_\phi(f V)_t = \dot{f}(t) V(t) + f(t) Der_\phi V(t);$$

3) for every $V \in Diff_m(\phi)$ and for every extension $\overline{V}$ of $V$ we have that

$$Der_\phi V(t) = \sum_{i=1}^{n} \dot{\phi}_i(t) \nabla_{E_i} \overline{V}(\phi(t)).$$

**Theorem.** *There exists one and only one covariant derivative $Der_\phi$ along $\phi$ with respect to $\nabla$. If $V = \sum_{j=1}^{m} V_j E'_j$ is an element of $Diff_m(\phi)$, then*

$$Der_\phi V(t) = \sum_{k=1}^{m}\left[\dot{V}_k(t) + \sum_{i=1}^{n}\sum_{j=1}^{m}\dot{\phi}_i(t)\Gamma_{ij}^k(\phi(t))V_j(t)\right]e'_k.$$

**Proof** By using property 1 of covariant derivatives we have that

$$Der_\phi V(t) = Der_\phi\left(\sum_{j=1}^{m} V_j E'_j\right)_t = \sum_{j=1}^{m} Der(V_j E'_j)_t.$$

By property 2 we have that

$$Der_\phi(V_j E'_j)_t = \dot{V}_j(t)E'_j(t) + V_j(t)Der_\phi(E'_j)_t = \dot{V}_j(t)e'_j + V_j(t)Der_\phi(E'_j)_t.$$

Since $\overline{E}'_j$ is an extension of $E'_j$, we have by property 3 that

$$Der_\phi(E'_j)_t = \sum_{i=1}^{n}\dot{\phi}_i(t)\nabla_{E_i}\overline{E}'_j(\phi(t)) = \sum_{i=1}^{n}\dot{\phi}_i(t)\sum_{k=1}^{m}\Gamma_{ij}^k(\phi(t))\overline{E}'_k(\phi(t)) = \sum_{i=1}^{n}\sum_{k=1}^{m}\dot{\phi}_i(t)\Gamma_{ij}^k(\phi(t))e'_k$$

This implies that $Der_\phi(V)_t$ must be equal to

$$\sum_{j=1}^{m}\left[\dot{V}_j(t)e'_j + V_j(t)\sum_{i=1}^{n}\sum_{k=1}^{m}\dot{\phi}_i(t)\Gamma_{ij}^k(\phi(t))e'_k\right] = \sum_{k=1}^{n}\dot{V}_k(t)e'_k + \sum_{i=1}^{n}\sum_{j,k=1}^{m}\dot{\phi}_i(t)\Gamma_{ij}^k(\phi(t))V_j(t)e'_k$$

which proves its existence and uniqueness. ∎

**Remark** The preceding equation can be written in the following matrix form

$$Der_\phi V(t) = \sum_{k=1}^{m}\left(\dot{V}_k(t) + \dot{\phi}(t)\Gamma^k(\phi(t))V(t)\right)e'_k.$$

## 2.5 PARALLEL TRANSPORT

A differential function on a curve $\phi$ whose covariant derivative along that curve is zero is called *parallel along* $\phi$. Since the $e_k$ are linearly independent, $V$ is parallel along $\phi$ if and only if it verifies the following linear and homogeneous system of ordinary differential equations,

$$\dot{V}_k(t) + \sum_{i=1}^{n}\sum_{j=1}^{m} \dot{\phi}_i(t)\, \Gamma_{ij}^{k}(\phi(t))\, V_j(t) = 0,$$

for $k = 1, \cdots, m$, which can be written in the matrix form

$$\dot{V}_k(t) + \dot{\phi}(t)\, \Gamma^{k}(\phi(t))\, V(t) = 0$$

for $k = 1, \cdots, m$.

We have the following result:

**Theorem.** Let $\nabla$ be a $(n,m)$-*connection, let* $\phi : I \to I\!R^n$ *be a curve, let* $t_0 \in I$ *and* $v \in I\!R^n$. *Then there exists a unique differentiable function* $V$ *parallel along* $\phi$ *with respect to* $\nabla$ *and such that* $V(t_0) = v$.

**Proof** See Theorem 4.11 of Lee (1997). ∎

Let $m = 1$; then the Christoffel symbols are $\Gamma_{i1}^{1}(x)$ for $i = 1, \cdots, n$, i.e., the Christoffel symbols form a vector with $n$ components. If we denote it $\Gamma(x)$ the equation for the parallel transport is

$$\dot{V}(t) = \left[-\dot{\phi}(t)\cdot \Gamma(\phi(t))\right] V(t) = \left(-\sum_{i=1}^{n} \dot{\phi}_i(t)\, \Gamma_{i1}^{1}(\phi(t))\right) V(t)$$

which admits the solution

$$V(t) = c\exp\left(-\int \dot{\phi}(t)\cdot \Gamma(\phi(t))\, dt\right) = c\exp\left(-\sum_{i=1}^{n}\int \dot{\phi}_i(t)\, \Gamma_{i1}^{1}(\phi(t))\, dt\right)$$

for any real constant $c$.

## 2.6 GEODESICS

Let $m = n$. The *acceleration of* $\phi$ *with respect to* $\nabla$ is $Der_\phi(\dot{\phi})$. We have that $Der_\phi(\dot{\phi})$ is equal to

$$\sum_{k=1}^{n}\left[\ddot{\phi}_k(t)+\sum_{i,j=1}^{n}\dot{\phi}_i(t)\dot{\phi}_j(t)\Gamma_{ij}^k(\phi(t))\right]e_k = \sum_{k=1}^{n}\left(\ddot{\phi}_k(t)+\dot{\phi}(t)\Gamma^k(\phi(t))\dot{\phi}(t)\right)e_k.$$

We say that $\phi$ is a *geodesic* if its acceleration is zero. This means that a curve is a geodesic if and only if its derivative is parallel along the curve. The equations for a geodesic are the following

$$\ddot{\phi}_k(t)+\sum_{i,j=1}^{n}\dot{\phi}_i(t)\dot{\phi}_j(t)\Gamma_{ij}^k(\phi(t))=0$$

for $k=1,\cdots,n$, or in the matrix form

$$\ddot{\phi}_k(t)+\dot{\phi}(t)\Gamma^k(\phi(t))\dot{\phi}(t)=0$$

for $k=1,\cdots,n$.

A geodesic is then the solution of a system of $k$ differential equations given above. It is natural to ask whether there is existence and uniqueness of solutions to such differential equations. This question is answered in the following:

**Theorem.** *Let $\nabla$ be a connection on $IR^n$ and $p,v \in IR^n$. Then there exists an open interval $I$ of $IR$, an element $t_0 \in I$ and a geodesic $\phi: I \to IR^n$ such that $\phi(t_0)=p$ and $\dot{\phi}(t_0)=v$. Moreover two such geodesics agree on their common domain.*

**Proof** See Theorem 4.10 of Lee (1997). ∎

**Remark.** Let $\phi: I \to IR^n$ be a geodesic and suppose that there exists $t_0 \in I$ such that $\dot{\phi}(t_0)=0$. Then $\phi$ is constant, that is $\dot{\phi}(t)=0$ for every $t$. Indeed, let $p=\phi(t_0)$ and let $\psi: I \to IR^n$ be the function sending any $t \in I$ to $p$. Then $\psi$ is a geodesic and verifies the same initial conditions as $\phi$, i.e., $\psi(t_0)=p$ and $\dot{\psi}(t_0)=0$, thus by the preceding theorem, $\phi$ and $\psi$ do coincide. ∎

## 2.7 RIEMANNIAN STRUCTURES

Let $M$ be a subset of $IR^n$; a *Riemannian metric* on $M$ is a differentiable function

$$g : M \times IR^n \times IR^n \to IR$$

such that for every $x \in M$ the function

$$IR^n \times IR^n \to IR$$
$$(u,v) \to g(x,u,v)$$

is a *scalar product*, i.e., it is symmetric and positive definite.

Let us fix a basis $\{e_1, \cdots, e_n\}$ of $IR^n$ and for $x \in M$ let us denote $g_{ij}(x) := g(x, e_i, e_j)$. Then the matrix $g(x) := \left(g_{ij}(x)\right)_{i,j=1,\cdots,n}$ is symmetric and positive definite. This implies that $g(x)$ is invertible and we denote $g^{ij}(x)$ the element of $i$-th row and $j$-th column of $g^{-1}(x)$, that is $g^{-1}(x) = \left(g^{ij}(x)\right)_{i,j=1,\cdots,n}$.

**Example.** Let $g(x)$ be the identity matrix for every $x \in IR^n$. Then $g$ defines a Riemannian structure on $IR^n$ which is called the *Euclidean metric*. ∎

**Example.** Let $IH^n$ be the positive half-space of $IR^n$, that is

$$IH^n = \{(x_1, \cdots, x_n) \in IR^n : x_n > 0\}.$$

We define on $IH^n$ the Riemannian metric $g$ such that if $x = (x_1, \cdots, x_n)$ then $g_{ij}(x) = \dfrac{\delta_{ij}}{x_n^2}$, that is the matrix corresponding to the point $p$ is equal to the identity matrix divided by $y^2$. $IH^n$ with this Riemannian metric is called the *Poincaré half-space*. ∎

We will use the notation $\langle u, v \rangle_x$ for $g(x,u,v)$. If $Y$ and $Z$ are vector fields we will denote $\langle Y, Z \rangle$ the function sending a $x \in IR^n$ to $\langle Y(x), Z(x) \rangle_x$.

Let $\nabla$ be a connection for $IR^n$; we will say that $\nabla$ is *compatible with* $g$ if for every vector fields $X, Y$ and $Z$ and for every $x \in M$ we have

$$D_X \langle Y, Z \rangle(x) = \langle \nabla_X Y(x), Z(x) \rangle_x + \langle Y(x), \nabla_X Z(x) \rangle_x.$$

(we recall that $D_X \langle Y, Z \rangle(x)$ denotes the derivative at $x$ of $\langle Y, Z \rangle$ along the direction $X$).

We say that $\nabla$ is *symmetric* if for every vector fields $X$ and $Y$ we have

$$\nabla_X Y - \nabla_Y X = D_X Y - D_Y X.$$

We have the following:

**Theorem.** *Let $g$ be a Riemannian metric on a subset of $IR^n$; then there exists one and only one connection which is symmetric and compatible with $g$. The Christoffel symbols of this connection are given by the formula*

$$\Gamma_{ij}^k(x) = \frac{1}{2} \sum_{l=1}^n g^{kl}(x) \left( \partial_i g_{jl}(x) + \partial_j g_{li}(x) - \partial_l g_{ij}(x) \right).$$

**Proof** See Theorem 5.4 of Lee (1997). ∎

The connection of the preceding theorem is called the *Levi-Civita connection*.

**Example.** Let $g$ be the Euclidean metric on $IR^n$. Since $g(x)$ is constant, then its derivatives are zero and then the Christoffel symbols are zero. ∎

**Example.** Let us compute the Christoffel symbols of the Poincaré half-space $IH^n$. We have that

$$\Gamma_{ij}^k(x) = \frac{1}{2} \sum_{l=1}^n g^{kl}(x) \left( \partial_i g_{jl}(x) + \partial_j g_{li}(x) - \partial_l g_{ij}(x) \right) =$$

$$= \frac{1}{2} \sum_{l=1}^n \delta_{kl} y^2 \left( \partial_i \left( \frac{\delta_{jl}}{x_n^2} \right) + \partial_j \left( \frac{\delta_{li}}{x_n^2} \right) - \partial_l \left( \frac{\delta_{ij}}{x_n^2} \right) \right) = \frac{x_n^2}{2} \left( \partial_i \left( \frac{\delta_{jk}}{x_n^2} \right) + \partial_j \left( \frac{\delta_{ki}}{x_n^2} \right) - \partial_k \left( \frac{\delta_{ij}}{x_n^2} \right) \right) =$$

$$= \frac{x_n^2}{2}\left(-\frac{2\delta_{jk}\delta_{ni}}{x_n^3} - \frac{2\delta_{ki}\delta_{nj}}{x_n^3} + \frac{2\delta_{ij}\delta_{nk}}{x_n^3}\right) = \frac{\delta_{ij}\delta_{nk} - \delta_{jk}\delta_{ni} - \delta_{ki}\delta_{nj}}{x_n}.$$

This implies that $\Gamma_{ij}^k(x) = 0$ if $n \neq i, j, k$ or if $i, j$ and $k$ are distinct two by two. The non-zero Christoffel symbols are thus:

$$\Gamma_{nn}^n(x) = -\frac{1}{x_n}; \; \Gamma_{ii}^n(x) = \frac{1}{x_n} \text{ for } i \neq n; \; \Gamma_{nk}^k(x) = -\frac{1}{x_n} \text{ for } k \neq n; \; \Gamma_{kn}^k(x) = -\frac{1}{x_n} \text{ for } k \neq n.$$

Now compute the matrices $\Gamma^k(x)$. We have that

$$\Gamma^n(x) = \frac{1}{x_n}\begin{pmatrix} 1 & 0 & \cdots & 0 & 0 \\ 0 & 1 & \cdots & 0 & 0 \\ \cdots & \cdots & \ddots & \cdots & \cdots \\ 0 & 0 & 0 & 1 & 0 \\ 0 & 0 & 0 & 0 & -1 \end{pmatrix},$$

that is $\Gamma^n(x)$ is $\dfrac{1}{x_n}$ times the diagonal matrix whose diagonal vector is $(1,1,\cdots,1,-1)$. For $k \neq n$ the matrix $\Gamma^k(x)$ is equal to $-\dfrac{1}{x_n}$ times the matrix whose $(k,n)$-th and $(n,k)$-th entries are 1 and the other are zero.

So the equations for the geodesics are the following:

$$\ddot{\phi}_n(t) + \frac{\left(\dot{\phi}_1(t)\right)^2 + \cdots \left(\dot{\phi}_{n-1}(t)\right)^2 - \left(\dot{\phi}_n(t)\right)^2}{\phi_n(t)} = 0,$$

$$\ddot{\phi}_k(t) - \frac{2\dot{\phi}_k(t)\dot{\phi}_n(t)}{\phi_n(t)} = 0 \text{ for } k = 1,\cdots,n-1.$$

## 2.8 CURVATURE

The *Riemann tensor* is defined as:

$$R^i_{jkl} = \partial_k \Gamma^i_{lj} - \partial_l \Gamma^i_{kj} + \sum_{r=1}^{n} \left( \Gamma^i_{kr} \Gamma^r_{lj} - \Gamma^i_{lr} \Gamma^r_{kj} \right);$$

the *Ricci tensor* is

$$R_{jl} = \sum_{i=1}^{n} R^i_{jil};$$

the *scalar curvature* is

$$R = \sum_{j,l=1}^{n} g^{jl} R_{jl}.$$

**Example.** For the Euclidean metric all the three functions defined above are zero since the Christoffel symbols are zero. ∎

**Example.** The non-zero components of the Riemann tensor for the Poincaré half-plane $IH^2$ are

$$R^1_{111}(x,y) = -\frac{1}{y^2}; \quad R^1_{122}(x,y) = -\frac{1}{y^2}; \quad R^1_{211}(x,y) = -\frac{1}{y^2};$$

$$R^1_{221}(x,y) = \frac{2}{y^2}; \quad R^2_{121}(x,y) = -\frac{1}{y^2}; \quad R^2_{211}(x,y) = \frac{1}{y^2}.$$

The components of the Ricci tensor are

$$R_{11}(x,y) = -\frac{2}{y^2}; \quad R_{12}(x,y) = R_{22}(x,y) = 0; \quad R_{21}(x,y) = -\frac{1}{y^2}.$$

Finally the scalar curvature is

$$R(x,y) = -\frac{2}{y}.$$

## 2.9 CURVES

An *admissible curve* is a function $\phi:[a,b] \to R^n$ which has finite right and left derivatives at all points and is differentiable with non-zero derivative unless at most finitely many points. By Remark (2.6), a non constant geodesic is an admissible curve since its derivative is always non-zero.

Let $[c,d]$ be an interval of real numbers and let $f:[c,d] \to [a,b]$ be a differentiable and invertible function whose inverse is also differentiable (in particular $f$ is monotonic). Then $\phi \circ f$ is called a *reparametrization* of $\phi$.

Let be given a Riemannian structure $g$ on $IR^n$. We call *speed of $\phi$ (with respect to $g$)* the function sending $t \in [a,b]$ to $\sqrt{\langle \dot\phi(t), \dot\phi(t) \rangle_{\phi(t)}}$. In particular we say that an admissible curve has *unit speed* if its speed is constantly equal to 1.

The *length of $\phi$ (with respect to $g$)* is defined as

$$L(\phi) := \int_a^b \sqrt{\langle \dot\phi(t), \dot\phi(t) \rangle_{\phi(t)}}\, dt = \int_a^b \sqrt{\dot\phi(t)\, g(\phi(t))\, \dot\phi(t)}\, dt = \int_a^b \sqrt{\sum_{i,j=1}^n g_{ij}(\phi(t))\, \dot\phi_i(t)\, \dot\phi_j(t)}\, dt.$$

We have the following

**Theorem.** *Let $\phi:[a,b] \to R^n$ be an admissible curve and $\psi$ a reparametrization of $\phi$; then $L(\phi) = L(\psi)$.*

**Proof** There exists an interval $[c,d]$ and a differentiable and invertible function $f:[c,d] \to [a,b]$ whose inverse is also differentiable and such that $\psi = \phi \circ f$. We have that $\dot\psi(t) = \dot\phi(f(t))\dot f(t)$, then

$$L(\psi) = \int_c^d \sqrt{\dot\psi(t)\, g(\psi(t))\, \dot\psi(t)}\, dt = \int_c^d \sqrt{\dot\phi(f(t))\, g(\phi \circ f(t))\, \dot\phi(f(t))}\, \big|\dot f(t)\big|\, dt =$$

$$= \int_a^b \sqrt{\dot\phi(t)\, g(\phi(t))\, \dot\phi(t)}\, dt = L(\phi),$$

where the penultimate equality follows from the rules of integration by substitution. ∎

We also have

**Proposition.** *Let $\phi:[a,b] \to R^n$ be an admissible curve and let $\lambda$ be the length of $\phi$. Let $f$ be the inverse of the function sending $t \in [a,b]$ to the length of $\phi$ restricted to $[a,t]$. Then $\psi := \phi \circ f$ is a unit speed reparametrization of $\phi$. Moreover if $h:[0,\lambda] \to [a,b]$ is a function such that $h(0) = a$, $h(\lambda) = b$ and the reparametrization $\phi \circ h$ has unit speed, then $h = f$.*

    **Proof** See Exercise 6.2 of Lee (1997). ∎

The last theorem tells that any admissible curve admits one and only one parametrization with unit speed. This parametrization is called *the parametrization by arc length*. For geodesics we have a stronger result:

**Theorem.** *Let $\phi:[a,b] \to R^n$ be a geodesic; then the function sending $t \in [a,b]$ to $\langle \dot\phi(t), \dot\phi(t) \rangle_{\phi(t)}$ is constant.*

    **Proof** See Lemma 5.5 of Lee (1997). ∎

    The last result implies that if $\phi:[a,b] \to R^n$ is a geodesic then there exists a constant $k$ such that $L(\phi) = k(b-a)$.

**Remark.** Let $\phi:[a,b] \to R^n$ be a non-constant geodesic and let $\psi = \phi \circ f$ be a reparametrization of $\phi$; we want to determine whether $\psi$ is a geodesic too. We have that $\psi_k = \phi_k \circ f$ and

$$\dot\psi_k(t) = \dot\phi_k(f(t))\dot f(t), \quad \dot\psi(t) = \dot\phi(f(t))\dot f(t), \quad \ddot\psi_k(t) = \ddot\phi_k(f(t))(\dot f(t))^2 + \dot\phi_k(f(t))\ddot f(t).$$

Thus

$$\ddot{\psi}_k(t) + \dot{\psi}(t)\Gamma^k(\psi(t))\dot{\psi}(t) =$$

$$= \ddot{\phi}_k(f(t))(\dot{f}(t))^2 + \dot{\phi}_k(f(t))\ddot{f}(t) + \dot{\phi}(f(t))\Gamma^k(\phi(f(t)))\dot{\phi}(f(t))(\dot{f}(t))^2 =$$

$$= \left[\ddot{\phi}_k(f(t)) + \dot{\phi}(f(t))\Gamma^k(\phi(f(t)))\dot{\phi}(f(t))\right](\dot{f}(t))^2 + \dot{\phi}_k(f(t))\ddot{f}(t) =$$

$$= \dot{\phi}_k(f(t))\ddot{f}(t),$$

where the last equality follows from the fact that $\phi$ is a geodesic. We have that $\dot{\phi}_k(f(t)) \neq 0$, since we have supposed that $\phi$ is non-constant; this implies that $\psi$ is a geodesic too if and only if $\ddot{f}(t) = 0$ for every $t$, that is if and only if there exist constants $\alpha$ and $\beta$ such that $f(t) = \alpha t + \beta$. ∎

Let $\phi:[a,b] \to R^n$ be an admissible curve; we say that $\phi$ is a *minimizing curve* if for every admissible curve $\psi:[c,d] \to R^n$ with the same endpoints (i.e., $\{\phi(a), \phi(b)\} = \{\psi(c), \psi(d)\}$), we have $L(\phi) \leq L(\psi)$. We have the following:

**Theorem.**

1) *Given two points $x$ and $y$ of $M$ there exists a minimizing curve whose endpoints are $x$ and $y$.*

2) *A minimizing curve with unit speed parametrization is a geodesic with respect to the Levi-Civita connection.*

**Proof** See Lemma 6.2 and Theorem 6.6 of Lee (1997). ∎

Given two points $x$ and $y$ of $M$, we call *geodesic distance between $x$ and $y$* the length of a minimizing curve joining them.

## 2.10 HEAT KERNEL ON A RIEMANNIAN STRUCTURE

Let $b^1, \cdots, b^n$ and $g^{ij}$ for $i, j \in \{1, \cdots, n\}$ be functions from $M$ to $I\!R$ and consider the following differential operator of order 2:

$$D\phi(x) := \sum_{i=1}^{n} b^i(x) \frac{\partial \phi}{\partial x_i}(x) + \sum_{i,j=1}^{n} g^{ij}(x) \frac{\partial^2 \phi}{\partial x_i \partial x_j}(x).$$

This operator is elliptic if and only if for any non-zero $x = (x_1, \cdots, x_n) \in M$ we have

$$\sum_{i,j=1}^{n} x_i x_j g^{ij}(x) \neq 0,$$

that is if and only if the matrix $g^{-1}(x)$ (and thus also the matrix $g(x)$) is positive definite for any non-zero $x$. This implies that if $g$ is a Riemannian metric on $M$ then the operator defined above is elliptic.

Now let $p = p(t, x, y)$ be a function from $I\!R \times I\!R^n \times I\!R^n$ to $I\!R$ and consider the equation

$$\frac{\partial p}{\partial t}(t, x, y) = Dp(t, x, y)$$

(where $D$ operates on $p$ considered as a function on $x$); it is called a *heat kernel equation*.

The function $p$ is called a *heat kernel* for the heat kernel equation defined above if $p$ is a solution to this equation and verifies also the following conditions:

1) it is of class $C^1$ with respect to $t$;

2) it is of class $C^2$ with respect to $x$;

3) verifies $\lim_{t \to 0} p(t, x, y) = \delta(x - y)$.

For $x, y \in M$ let us denote $dist_g(x, y)$ the geodesic distance between $x$ and $y$ with respect to the Riemannian metric $g$. We define the *Synge world function with respect to $g$* as

$$Syn_g(x, y) := \frac{1}{2} \bigl( dist_g(x, y) \bigr)^2.$$

Now let us denote $C(x, y)$ the $n \times n$ matrix whose $(i, j)$ – entry for $i, j \in \{1, \cdots, n\}$ is

$$-\frac{\partial^2 Syn_g(x, y)}{\partial x_i \partial y_j};$$

we define the *Van Vleck-Morette determinant with respect to $g$* as

$$Van_g(x, y) := \frac{\det(C(x, y))}{\sqrt{\det g(x) \det g(y)}}.$$

Now for $i = 1, \cdots, n$ let $A_i : M \to \mathbb{R}$ be defined by

$$A_i(x) := \frac{1}{2} \sum_{h=1}^{n} g_{ih}(x) \left[ b^h(x) - \frac{\sum_{k=1}^{n} \partial_k \left( \sqrt{\det g(x)} \, g^{hk}(x) \right)}{\sqrt{\det g(x)}} \right]$$

and consider the $(n,1)$ – connection whose Christoffel symbols are $A_1, \cdots, A_n$. Let $\phi = (\phi_1, \cdots, \phi_n)$ be a curve; the parallel transport to $\phi$ with respect to this connection is given by (see also 3.135 of Avramidi)

$$Par_\phi(t) = \exp\left( -\sum_{i=1}^{n} \int_0^t \dot{\phi}_i(t) A_i(\phi(t)) du \right).$$

We have the following crucial

**Theorem.** *Let $p = p(t, x, y)$ be the heat kernel for the heat kernel equation given above. Let us fix $x, y \in M$ and let $\phi : [0, t] \to M$ be a minimizing curve such that $\phi(0) = x$ and $\phi(t) = y$. Then there exist functions $a_k(x, y)$ for $k \in \mathbb{N}$ such that $p$ admits the following asymptotic expansion for $t \to 0$,*

$$p(t, x, y) = \sqrt{\frac{\det g(y) \, Van_g(x, y)}{(4\pi t)^n}} \, Par_\phi(t) \exp\left( -\frac{Syn_g(x, y)}{2t} \right) \sum_{k=0}^{+\infty} a_k(x, y) t^k.$$

**Proof**  See [H-L 2008] and [Avramidi]. ∎

We have that $a_0(x_1, y_1; x_2, y_2) = 1$ and that

$$a_1(x, y; x, y) = \frac{1}{6} R(x, y) + Q(x, y),$$

where $R(x, y)$ is the scalar curvature.

## 2.11 ELLIPTIC OPERATORS ON A RIEMANNIAN STRUCTURE

Let $g$ be a Riemannian metric on $M$; our goal in this section is to define a Riemannian analogue of the notion of Laplacian. For that first we will define Riemannian analogues of the gradient and of the divergence.

Let $x \in M$, let $\phi: M \to R$ be a function admitting partial derivatives at $x$ and let

$$grad\ \phi(x) = \sum_{i=1}^{n} (\partial_i \phi(x)) e_i$$

be the gradient of $\phi$ at $x$ (we recall that $\{e_1, \cdots, e_n\}$ is the canonical basis of $IR^n$). The *gradient of* $\phi$ *with respect to* $g$ *at* $x$ is defined as

$$grad_g\ \phi(x) = g^{-1}(x)\ grad\ \phi(x),$$

i.e.,

$$grad_g\ \phi(x) = \sum_{i,j=1}^{n} (g^{ij}(x) \partial_j \phi(x)) e_i.$$

In particular the standard gradient is the one corresponding to the Euclidean metric.

Let $\phi_1, \cdots, \phi_n: M \to IR$ be such that $\phi_i$ admits $i$-th partial derivative at $x$ and let $\Phi = (\phi_1, \cdots, \phi_n)$ be the vector function whose components are the $\phi_i$. The *divergence of* $\Phi$ *with respect to* $g$ *at* $x$ is defined as

$$div_g\ \Phi(x) = \frac{1}{\sqrt{\det(g(x))}}\ div\left(\sqrt{\det(g(x))}\ \Phi(x)\right),$$

i.e.,

$$div_g \, \Phi(x) = \frac{1}{\sqrt{\det(g(x))}} \sum_{i=1}^{n} \partial_i \left( \sqrt{\det(g(x))} \, \phi_i(x) \right).$$

As for the gradient, the standard divergence is the divergence with respect to the Euclidean metric.

Let $\phi : M \to R$ be a function admitting partial second derivatives at $x$; the *Laplace-Beltrami operator of* $\phi$ *at* $x$ *with respect to* $g$ is defined as the $g$ – divergence of the $g$ – gradient of $\phi$ at $x$, that is as

$$\Delta_g \phi(x) := div_g \left( grad_g \, \phi(x) \right),$$

that is

$$\Delta_g \phi(x) = \frac{1}{\sqrt{\det(g(x))}} div \left( \sqrt{\det(g(x))} \, grad_g \, \phi(x) \right) =$$

$$= \frac{1}{\sqrt{\det(g(x))}} div \left( \sqrt{\det(g(x))} \, g^{-1}(x) \, grad \, \phi(x) \right) =$$

$$= \frac{1}{\sqrt{\det(g(x))}} \sum_{i,j=1}^{n} \partial_i \left( \sqrt{\det(g(x))} \, g^{ij}(x) \partial_j \phi(x) \right),$$

where as usual $g^{ij}(x)$ denotes the $(i, j)$ – th element of $g^{-1}(x)$.

Now let $D$ be the differential operator defined by

$$D\phi(x) := \sum_{i=1}^{n} b^i(x) \frac{\partial \phi}{\partial x_i}(x) + \sum_{i,j=1}^{n} g^{ij}(x) \frac{\partial^2 \phi}{\partial x_i \partial x_j}(x).$$

For $D$ the elliptic operator defined as above and let $Q : M \to \mathit{IR}$ be defined by

$$Q(x) := \sum_{i,j=1}^{n} \left[ g^{ij}(x) A_i(x) A_j(x) + \frac{\partial_i \left( \sqrt{\det g(x)} \, g^{ij}(x) A_j(x) \right)}{\sqrt{\det g(x)}} \right],$$

where the $A_i$ were defined above. Then it can be proved that

$$D\phi(x) = \left( \frac{\sum_{i,j=1}^{n} \left[ (\partial_i + A_i(x)) \sqrt{\det g(x)} g^{ij}(x) (\partial_j + A_j(x)) \right]}{\sqrt{\det g(x)}} + Q(x) \right) \phi(x).$$

We observe that this expression is equal to the Laplace-Beltrami operator if and only if $A_i = 0$ for every $i$.

## 2.12 CHANGE OF COORDINATES IN A RIEMANNIAN STRUCTURE

Let $f : I\!R^n \to I\!R^n$ be a bijective and differentiable function such that $f(M) = M$ and as usual let us denote $f_1, \cdots, f_n$ the components of $f$. Let $x \in I\!R^n$; the *Jacobian of $f$ at $x$*, denoted $Jac\, f(x)$ is the square matrix of order $n$ whose $(i, j)$-element is $\frac{\partial f_i}{\partial x_j}(x)$.

Let $\phi : [a, b] \to M$ be a curve and let us consider the curve $f \circ \phi$. The components of $f \circ \phi$ are $f_1 \circ \phi, \cdots, f_n \circ \phi$ and the derivative of each component is

$$\frac{d}{dt} f_i \circ \phi(t) = grad\, f_i(\phi(t)) \cdot \dot\phi(t) = \sum_{h=1}^{n} \partial_h f_i(\phi(t)) \dot\phi_h(t),$$

so the length of $f \circ \phi$ with respect to $g$ is equal to

$$L_g(f \circ \phi) = \int_a^b \sqrt{\sum_{i,j=1}^{n} g_{ij}(f \circ \phi(t)) \sum_{h=1}^{n} \partial_h f_i(\phi(t)) \dot\phi_h(t) \sum_{k=1}^{n} \partial_k f_j(\phi(t)) \dot\phi_k(t)}\, dt =$$

$$= \int_a^b \sqrt{\sum_{h,k=1}^{n} \left( \sum_{i,j=1}^{n} \partial_h f_i(\phi(t)) g_{ij}(f \circ \phi(t)) \partial_k f_j(\phi(t)) \right) \dot\phi_h(t) \dot\phi_k(t)}\, dt.$$

Let us denote $f * g(x)$ the matrix function whose $(h, k)$-element is

$$\sum_{i,j=1}^{n} \partial_h f_i(x) g_{ij}(f(x)) \partial_k f_j(x),$$

that is the product of the $h$-th column of $Jac\, f(x)$, times $g(x)$, times the $k$-th column of $Jac\, f(x)$. Then we have that

$$L_g(f \circ \phi) = \int_a^b \sqrt{\sum_{h,k=1}^n f * g(\phi(t))\dot{\phi}_h(t)\dot{\phi}_k(t)}\, dt = L_{f*g}(\phi),$$

that is the length of $f \circ \phi$ with respect to $g$ is equal to the length of $\phi$ with respect to $f * g$.

Thus we can say that if we make a change of coordinates in $M$ by means of $f$ then the metric changes from $g$ to $f * g$.

As observed above, $f * g(x)$ is the product of the transpose of $Jac\, f(x)$, times $g(x)$, times $Jac\, f(x)$, that is

$$f * g(x) = {}^t Jac\, f(x)\, g(x)\, Jac\, f(x).$$

Suppose that we change the coordinates from $x_1,\cdots,x_n$ to $f_1,\cdots,f_n$. This means that $\phi$ is changed to $f^{-1} \circ \phi$. Let $L(x) := Jac\, f(x)$; since $Jac\, f^{-1}(x) = L(x)^{-1}$, we have that the metric becomes

$${}^t L^{-1}(x)\, g(x)\, L^{-1}(x)$$

and its inverse is

$$L(x)\, g^{-1}(x)\, {}^t L(x).$$

## 2.13 CHAIN RULE FOR PARTIAL DERIVATIVES

Let $M \subset R^n$, let $p : M \to I\!R$ be a twice differentiable function and let $f : M \to M$ be a twice differentiable and bijective function. We denote $f_1,\cdots,f_n$ the components of $f$. Let $x_1,\cdots,x_n$ be real variables and suppose that $p$ is given as a function of $x_1,\cdots,x_n$, that is $p = p(x_1,\cdots,x_n)$. Let also $x := (x_1,\cdots,x_n)$.

We can consider $f_1,\cdots,f_n$ as real variables and express $p$ as a function of $f_1,\cdots,f_n$, that is $p = p(f_1,\cdots,f_n)$. By the *chain rule of partial derivatives*, we have that for $j = 1,\cdots,n$,

$$\frac{\partial p}{\partial x_j}(x) = \sum_{k=1}^{n} \frac{\partial p}{\partial f_k}(f(x)) \frac{\partial f_k}{\partial x_j}(x),$$

therefore

$$\frac{\partial^2 p}{\partial x_i \partial x_j} = \sum_{k=1}^{n} \frac{\partial}{\partial x_i}\left(\frac{\partial p}{\partial f_k} \frac{\partial f_k}{\partial x_j}\right).$$

We have that

$$\frac{\partial}{\partial x_i}\left(\frac{\partial p}{\partial f_k} \frac{\partial f_k}{\partial x_j}\right) = \frac{\partial}{\partial x_i}\left(\frac{\partial p}{\partial f_k}\right)\frac{\partial f_k}{\partial x_j} + \frac{\partial p}{\partial f_k}\frac{\partial^2 f_k}{\partial x_i \partial x_j}$$

and that

$$\frac{\partial}{\partial x_i}\left(\frac{\partial p}{\partial f_k}\right) = \sum_{h=1}^{n} \frac{\partial^2 p}{\partial f_h \partial f_k}\frac{\partial f_h}{\partial x_i},$$

so we have

$$\frac{\partial^2 p}{\partial x_i \partial x_j}(x) = \sum_{k=1}^{n}\left[\frac{\partial}{\partial x_i}\left(\frac{\partial p}{\partial f_k}\right)\frac{\partial f_k}{\partial x_j} + \frac{\partial p}{\partial f_k}\frac{\partial^2 f_k}{\partial x_i \partial x_j}\right] = \sum_{k=1}^{n}\left[\sum_{h=1}^{n}\frac{\partial^2 p}{\partial f_h \partial f_k}\frac{\partial f_h}{\partial x_i}\frac{\partial f_k}{\partial x_j} + \frac{\partial p}{\partial f_k}\frac{\partial^2 f_k}{\partial x_i \partial x_j}\right] =$$

$$= \sum_{h,k=1}^{n}\frac{\partial^2 p}{\partial f_h \partial f_k}(f(x))\frac{\partial f_h}{\partial x_i}(x)\frac{\partial f_k}{\partial x_j}(x) + \sum_{k=1}^{n}\frac{\partial p}{\partial f_k}(f(x))\frac{\partial^2 f_k}{\partial x_i \partial x_j}(x)$$

## 2.14 CHANGE OF COORDINATES IN THE KOLMOGOROFF BACKWARD EQUATION

Let $n$ be a natural number and let $p : IR \times M \times M \to IR$, let $D$ be the differential operator defined by

$$D\phi(x) := \sum_{i=1}^{n} b^i(x)\frac{\partial \phi}{\partial x_i}(x) + \sum_{i,j=1}^{n} g^{ij}(x)\frac{\partial^2 \phi}{\partial x_i \partial x_j}(x)$$

and let $p$ verify the partial differential equation

$$\frac{\partial p}{\partial t}(t, x, y) = Dp(t, x, y),$$

where $t \in \mathbb{R}$, $x, y \in M$ and $D$ acts on $p$ as a function of $x$. Let $f : M \to M$ be a twice differentiable and bijective function whose second partial derivatives are zero and let us consider $p$ as a function of $t, f$ and $y$. By the result given in the preceding paragraph we have that

$$\frac{\partial p}{\partial x_i}(x) = \sum_{h=1}^{n} \frac{\partial p}{\partial f_h}(f(x)) \frac{\partial f_h}{\partial x_i}(x)$$

and that

$$\frac{\partial^2 p}{\partial x_i \partial x_j}(x) = \sum_{h,k=1}^{n} \frac{\partial^2 p}{\partial f_h \partial f_k}(f(x)) \frac{\partial f_h}{\partial x_i}(x) \frac{\partial f_k}{\partial x_j}(x).$$

Thus

$$Dp(t, f(x), y) = \sum_{i=1}^{n} b^i(x) \sum_{h=1}^{n} \frac{\partial p}{\partial f_h}(f(x)) \frac{\partial f_h}{\partial x_i}(x) + \sum_{i,j=1}^{n} g^{ij}(x) \sum_{h,k=1}^{n} \frac{\partial^2 p}{\partial f_h \partial f_k}(f(x)) \frac{\partial f_h}{\partial x_i}(x) \frac{\partial f_k}{\partial x_j}(x) =$$

$$= \sum_{h=1}^{n} \left( \sum_{i=1}^{n} \frac{\partial f_h}{\partial x_i}(x) b^i(x) \right) \frac{\partial p}{\partial f_h}(f(x)) + \sum_{h,k=1}^{n} \left( \sum_{i,j=1}^{n} \frac{\partial f_h}{\partial x_i}(x) g^{ij}(x) \frac{\partial f_k}{\partial x_j}(x) \right) \frac{\partial^2 p}{\partial f_h \partial f_k}(f(x)).$$

Let $L(x)$ be the Jacobian matrix at $x$ of $f$ with respect to the variables $x_1, \cdots, x_n$, that is

$$L(x) = \left( \frac{\partial f_i}{\partial x_j}(x) \right)_{i,j=1,\cdots,n}.$$

By elementary rules of the product of matrices we have that

$$\sum_{i,j=1}^{n} \frac{\partial f_h}{\partial x_i}(x) g^{ij}(x) \frac{\partial f_k}{\partial x_j}(x)$$

is the $(h, k)$ – entry of the matrix

$$L(x) g^{-1}(x) {}^t L(x).$$

We observe that since $f$ is bijective, we can consider $x$ as a function of $f$, $x = x(f)$, thus setting

$$\gamma^{-1}(f) := L(x(f))g^{-1}(x(f))\,{}^tL(x(f))$$

and

$$\beta^h(f) := \sum_{i=1}^{n} \frac{\partial f_h}{\partial x_i}(x(f))b^i(x(f)) = b(x(f)) \cdot \mathrm{grad}\, f_h(x(f)),$$

we have that

$$Dp(t,f,y) = \sum_{h=1}^{n} \beta^h(f)\frac{\partial p}{\partial f_h}(x(f)) + \sum_{h,k=1}^{n} \gamma^{hk}(x(f))\frac{\partial^2 p}{\partial f_h \partial f_k}(x(f)).$$

Since the derivative of $p$ with respect to $t$ does not change when considering $p$ as a function $t, f$ and $y$ we have that

$$\frac{\partial p}{\partial t}(t,f,y) = \sum_{h=1}^{n} \beta^h(f)\frac{\partial p}{\partial f_h}(x(f)) + \sum_{h,k=1}^{n} \gamma^{hk}(x(f))\frac{\partial^2 p}{\partial f_h \partial f_k}(x(f)).$$

# 3 The Poincaré half-plane

In this chapter we study the *Poincaré half-plane*, a Riemannian metric on the positive half-plane which will result to be the metric for the SABR model.

The chapter is organized as follows: after introducing the Poincaré model for a generic dimension $n$ and computed the equations for the geodesics, we study in details the case $n=2$. We find two classes of solutions to these equations, namely the vertical lines and the positive semicircles with center on the $x-$axis. We compute the geodesic distance for them and show that they are the only geodesics.

A vertical line has equation

$$\phi(t) = \left(a, b e^{\pm pt}\right),$$

for $a, b$ and $p$ real constants with $b > 0$; a positive semicircle has equation

$$\phi(t) = \left(r\tanh(qt) + c, \ \frac{r}{\cosh(qt)}\right).$$

## 3.1 POINCARÉ HALF-SPACE

Let $IH^n$ be the positive half-space of $IR^n$, that is

$$IH^n = \left\{(x_1, \cdots, x_n) \in IR^n : x_n > 0\right\}.$$

We define on $IH^n$ the Riemannian metric $g$ such that if $p = (x_1, \cdots, x_n)$ then $g_{ij}(p) = \dfrac{\delta_{ij}}{x_n^2}$, that is the matrix corresponding to the point $p$ is equal to the identity matrix divided by $y^2$. This means that the scalar product at $(x_1, \cdots, x_n)$ is equal to the standard Euclidean scalar product divided by $y^2$. Thus we have that $g^{ij}(p)$, the $(i,j)$-th element of the inverse of $g(p)$, is equal to $\delta_{ij} x_n^2$.

As seen before the equations for the geodesics are the following:

$$\ddot{\phi}_n(t) + \frac{\left(\dot{\phi}_1(t)\right)^2 + \cdots + \left(\dot{\phi}_{n-1}(t)\right)^2 - \left(\dot{\phi}_n(t)\right)^2}{\phi_n(t)} = 0,$$

$$\ddot{\phi}_k(t) - \frac{2\dot{\phi}_k(t)\dot{\phi}_n(t)}{\phi_n(t)} = 0 \quad \text{for } k = 1, \cdots, n-1.$$

Now let $n = 2$ and let us set $x := \phi_1$, $y := \phi_2$. Then the system of differential equations is

$$\begin{cases} \ddot{x} - \dfrac{2\dot{x}\dot{y}}{y} = 0 \\ \ddot{y} + \dfrac{(\dot{x})^2 - (\dot{y})^2}{y} = 0 \end{cases},$$

which is equivalent to

$$\begin{cases} \ddot{x}y - 2\dot{x}\dot{y} = 0 \\ y\ddot{y} + (\dot{x})^2 - (\dot{y})^2 = 0 \end{cases}.$$

## 3.2 GEODESICS OF THE POINCARÉ HALF-PLANE: VERTICAL LINES

First suppose that $\dot{x}(t) = 0$ for every $t$; this means that there exists a constant $a$ such that $x(t) = a$ for every $t$. Moreover the second equation becomes

$$\ddot{y} - \frac{(\dot{y})^2}{y} = 0.$$

To solve it, let us observe that

$$\frac{d}{dt}\left(\frac{\dot{y}}{y}\right) = \frac{\ddot{y}y - (\dot{y})^2}{y^2} = \frac{1}{y}\left(\ddot{y} - \frac{(\dot{y})^2}{y}\right),$$

thus the second equation is verified if and only if $\frac{d}{dt}\left(\frac{\dot{y}}{y}\right) = 0$. Then there exists a constant $b$ such that $\frac{\dot{y}}{y} = b$ and therefore there is another constant $c$ such that $y = be^{ct}$. Since $y > 0$ then $b$ must be positive.

In conclusion the expression for the curve is $\phi(t) = (a, be^{ct})$, where $a, b$ and $c$ are real numbers. This curve is a vertical line.

Now we suppose that this curve is parametrized by arc length. By Proposition 2.9, we have that the speed of the curve is equal to 1. Thus since $\dot{\phi}(t) = (0, bc\, e^{ct})$ we have that

$$\sqrt{\langle \dot{\phi}(t), \dot{\phi}(t) \rangle_{\phi(t)}} = \sqrt{\frac{b^2 c^2 e^{2ct}}{b^2 e^{2ct}}} = \sqrt{c^2} = |c|,$$

that we set it equal to 1, so $c = \pm 1$. Thus the geodesics are $\phi(t) = (a, be^{\pm t})$ for constants $a$ and $b$.

## 3.3 GEODESIC DISTANCE FOR VERTICAL LINES

Let us take two points $(a, y_1)$ and $(a, y_2)$ with the same $x$-coordinate. Up to reparametrization, there is only one geodesic of the kind found above joining these two points, the straight line segment $x = a$ comprised between $y_1$ and $y_2$. We will prove that, supposing without loss of generality that $y_1 < y_2$, the length of the arc of curve joining them is equal to $\log \frac{y_2}{y_1}$.

Indeed, let us take that curve, which we call $\phi$ and let $t_1, t_2$ be such that $\phi(t_1) = (a, y_1)$ and $\phi(t_2) = (a, y_2)$. Since the curve is parametrized by arc length, the length of the arc of $\phi$ comprised between $\phi(t_1)$ and $\phi(t_2)$ is $|t_2 - t_1|$.

First let the geodesic be $\phi(t) = (a, be^t)$; then from $y_i = be^{t_i}$ we obtain that $t_i = \log \frac{y_i}{b}$ for $i = 1,2$. Since in this case $y(t)$ is increasing, from $y_1 < y_2$ we obtain that $t_1 < t_2$, thus

$$|t_2 - t_1| = t_2 - t_1 = \log \frac{y_2}{b} - \log \frac{y_1}{b} = \log \frac{y_2}{y_1}.$$

Now let the geodesic be $\phi(t) = (a, be^{-t})$; then in this case $t_i = -\log \frac{y_i}{b}$. The function $y(t)$ is decreasing, thus $y_1 < y_2$ implies that $t_2 < t_1$ and

$$|t_2 - t_1| = t_1 - t_2 = -\log \frac{y_1}{b} + \log \frac{y_2}{b} = \log \frac{y_2}{y_1}.$$

## 3.4 GEODESICS OF THE POINCARÉ HALF-PLANE: SEMICIRCLES

Now let $\dot{x}$ be non-necessarily zero. We observe that the first equation gives

$$\ddot{x} - \frac{2\dot{x}\dot{y}}{y} = 0 \quad \Leftrightarrow \quad \frac{\ddot{x}y - 2\dot{x}\dot{y}}{y} = 0 \quad \Leftrightarrow \quad \frac{\ddot{x}y - 2\dot{x}\dot{y}}{y^3} = 0 \quad \Leftrightarrow \quad D\left[\frac{\dot{x}}{y^2}\right] = 0,$$

which is true if and only if there exists a non-zero constant $r$ such that $\dot{x} = \frac{y^2}{r}$.

As before, we impose that $\langle \dot{\phi}(t), \dot{\phi}(t) \rangle_{\phi(t)} = 1$, that is

$$\frac{(\dot{x})^2 + (\dot{y})^2}{y^2} = 1 \quad \Leftrightarrow \quad (\dot{x})^2 + (\dot{y})^2 = y^2 \quad \Leftrightarrow \quad (\dot{y})^2 = y^2 - (\dot{x})^2 = y^2 - \frac{y^4}{r^2}.$$

Let $\dot{y} = y\sqrt{1 - \frac{y^2}{r^2}}$; this is a solution of the preceding equation. Set $\zeta := \frac{y}{r}$; then

$$\dot{\zeta} = \frac{\dot{y}}{r} = \frac{y}{r}\sqrt{1 - \frac{y^2}{r^2}} = \zeta\sqrt{1 - \zeta^2}.$$

We now recall the following result from calculus: if $f$ is an injective and differentiable real function, then $D[f^{-1}](t) = \dfrac{1}{D[f](f^{-1}(t))}$. Thus

$$\zeta^{-1}(z) = \int D[\zeta^{-1}](z)dz = \int \frac{1}{\dot\zeta(\zeta^{-1}(z))}dz = \int \frac{1}{\zeta(\zeta^{-1}(z))\sqrt{1-[\zeta(\zeta^{-1}(z))]^2}}dz = \int \frac{1}{z\sqrt{1-z^2}}dz.$$

Now set $\theta := \arccos z$, that is $z = \cos\theta$ and $dz = -\sin\theta\, d\theta$. Thus the preceding integral is equal to

$$-\int \frac{\sin\theta}{\cos\theta\sqrt{1-\cos^2\theta}}d\theta = -\int \frac{1}{\cos\theta}d\theta.$$

Observe that

$$D\left[\log\left(\frac{1+\sin\theta}{\cos\theta}\right)\right] = \frac{\cos\theta}{1+\sin\theta}\frac{\cos^2\theta+\sin^2\theta+\sin\theta}{\cos^2\theta} = \frac{1}{\cos\theta},$$

thus

$$-\int \frac{\sin\theta}{\cos\theta\sqrt{1-\cos^2\theta}}d\theta = -\log\left(\frac{1+\sin\theta}{\cos\theta}\right) = -\log\left(\frac{1+\sqrt{1-z^2}}{z}\right).$$

Therefore

$$t = \zeta^{-1}(\zeta(t)) = -\log\left(\frac{1+\sqrt{1-\zeta^2(t)}}{\zeta(t)}\right) \Leftrightarrow$$

$$\Leftrightarrow \quad e^{-t} = \frac{1+\sqrt{1-\zeta^2(t)}}{\zeta(t)} \quad \Leftrightarrow \quad e^{-t}\zeta(t)-1 = \sqrt{1-\zeta^2(t)} \quad \Leftrightarrow$$

$$\Leftrightarrow \quad e^{-2t}\zeta^2(t) - 2e^{-t}\zeta(t)+1 = 1-\zeta^2(t) \quad \Leftrightarrow$$

$$\Leftrightarrow \quad (1+e^{-2t})\zeta^2(t) - 2e^{-t}\zeta(t) = 0 \quad \Leftrightarrow$$

$$\Leftrightarrow \quad ((1+e^{-2t})\zeta(t) - 2e^{-t})\zeta(t) = 0 \quad \Leftrightarrow^2 \quad (1+e^{-2t})\zeta(t) - 2e^{-t} = 0 \quad \Leftrightarrow$$

---

[2] Since $\zeta = \alpha y$ then $\zeta > 0$

$$\Leftrightarrow \quad \zeta(t) = \frac{2e^{-t}}{1+e^{-2t}} = \frac{1}{\cosh t} \quad \Leftrightarrow \quad y(t) = \frac{r}{\cosh t}.$$

Therefore

$$\dot{x} = \frac{y^2}{r} = \frac{r}{\cosh^2 t}$$

and thus

$$x = r \tanh t + c$$

for some constant $c$. In conclusion the geodesic is

$$\phi(t) = \left( r \tanh t + c, \; \frac{r}{\cosh t} \right).$$

Since $y > 0$ then we have that $r > 0$.

We now observe that

$$(x(t) - c)^2 + (y(t))^2 = r^2 \left( \tanh^2 t + \frac{1}{\cosh^2 t} \right) = r^2,$$

since we have the equality

$$\tanh^2 t + \frac{1}{\cosh^2 t} = 1.$$

Since, moreover for $t \in ]-\infty, +\infty[$ the function $\tanh t$ assumes all the values in $]-1,1[$ and the function $1/\cosh t$ all the values in $]0,1[$, we have that the geodesic we have found is the positive semicircle with center $(c,0)$ and radius $r$.

## 3.5 GEODESIC DISTANCE FOR SEMICIRCLES

Let us take two points $(x_1, y_1)$ and $(x_2, y_2)$ with $x_1 \neq x_2$. First we show that up to reparametrization, there is only one semicircle joining these two points. This is equivalent to prove that there is only one point in the $x$-axis with the same distance from $(x_1, y_1)$ and $(x_2, y_2)$.

Indeed let $c$ be such that the distance from $(c,0)$ to $(x_1, y_1)$ is equal to that from $(c,0)$ to $(x_2, y_2)$; then the equality

$$\sqrt{(c-x_1)^2 + y_1^2} = \sqrt{(c-x_2)^2 + y_2^2}$$

has the unique solution

$$c = \frac{x_2^2 - x_1^2 + y_2^2 - y_1^2}{2(x_2 - x_1)}.$$

Now let us compute the geodesic distance of two points joined by a semicircle. We saw that

$$t = -\log\left(\frac{1+\sqrt{1-\zeta^2(t)}}{\zeta(t)}\right) = \log\left(\frac{y(t)}{r+\sqrt{r^2 - y^2(t)}}\right).$$

Let $\phi$ be the semicircle and let $t_1, t_2$ be such that $(x_1, y_1) = \phi(t_1)$ and $(x_2, y_2) = \phi(t_2)$. We suppose that $t_1 < t_2$, so the length of the arc of the geodesic joining $(x_1, y_1)$ and $(x_2, y_2)$ is

$$t_2 - t_1 = \log\left(\frac{y_2}{r+\sqrt{r^2 - y_2^2}}\right) - \log\left(\frac{y_1}{r+\sqrt{r^2 - y_1^2}}\right) = -\log\left(\frac{y_2}{y_1}\frac{r+\sqrt{r^2 - y_1^2}}{r+\sqrt{r^2 - y_2^2}}\right).$$

It can be proved that the latter expression is equal to

$$\operatorname{arccos} h\left(1 + \frac{(x_2 - x_1)^2 + (y_2 - y_1)^2}{2 y_1 y_2}\right).$$

## 3.6 PROOF OF THE UNIQUENESS OF THE GEODESICS

Now we prove that vertical lines and semicircles are the only geodesics for the Poincaré half-plane. For that we have to show that given real numbers $u_1, u_2, v_1, v_2$ there exists a curve $\phi$ which is a vertical line or a semicircle and a $t_0$ such that $\phi(t_0) = (u_1, u_2)$, $\dot{\phi}(t_0) = (v_1, v_2)$. We observe that since the $y$-coordinate is positive, we must have $u_2 > 0$.

By Remark (2.9), for every real number $\alpha$, the composition of a geodesic $\phi$ with the function $t \to \alpha t$ is still a geodesic, so the two classes of geodesics we have found, together with their derivatives, are

$$\phi(t) = \left(a, b e^{\pm \alpha t}\right), \quad \dot{\phi}(t) = \left(0, \pm b p e^{\pm \alpha t}\right)$$

with $a$ and $b$ arbitrary constants and

$$\psi(t) = \left(r \tanh(\alpha t) + c, \frac{r}{\cosh(\alpha t)}\right), \quad \dot{\psi}(t) = \left(\frac{pr}{\cosh^2(\alpha t)}, \frac{-pr \sinh(\alpha t)}{\cosh^2(\alpha t)}\right),$$

with $r > 0$ and $c$ constants.

Let $v_1 = 0$ and take $\phi(t) = (a, b e^{\alpha t})$; then

$$\phi(0) = (a, b), \quad \dot{\phi}(t) = (0, b\alpha)$$

and we have a solution for $a = u_1$, $b = u_2$, $\alpha = \frac{v_2}{u_2}$, that is the curve

$$\phi(t) := \left(u_1, u_2 e^{\frac{v_2}{u_2} t}\right)$$

is a geodesic and verifies the initial condition for $t_0 = 0$.

Now let $v_1 \neq 0$; we have to find constants $r, \alpha, c$ and $t_0$ (with $r > 0$) such that

$$r \tanh(\alpha t_0) + c = u_1, \quad \frac{r}{\cosh(\alpha t_0)} = u_2, \quad \frac{\alpha r}{\cosh^2(\alpha t_0)} = v_1, \quad \frac{-pr \sinh(\alpha t_0)}{\cosh^2(\alpha t_0)} = v_2.$$

After long (but simple) calculations we find that the solution to this system is given by

$$r = u_2 \sqrt{\frac{v_1^2 + v_2^2}{v_1^2}}, \quad \alpha = \frac{v_1}{u_2} \sqrt{\frac{v_1^2 + v_2^2}{v_1^2}}, \quad c = \frac{u_1 v_1 + u_2 v_2}{v_1}, \quad t_0 = \frac{u_2}{v_1} \sqrt{\frac{v_1^2}{v_1^2 + v_2^2}} \operatorname{arcsin} h\left(-\frac{v_2}{v_1}\right).$$

## 3.7 GEODESIC CONNECTING TWO GIVEN POINTS

Let us given two points $z_1 = (x_1, y_1)$ and $z_2 = (x_2, y_2)$ of the Poincaré half plane $IH^2$ and let us find the geodesic connecting them, that is a geodesic

$$\psi : [0, \tau] \to IH^2$$

for some $\tau > 0$ such that $\psi(0) = z_1$ and $\psi(\tau) = z_2$. First let us suppose that $x_1 \neq x_2$ and consider the semicircle

$$\psi(t) = \left( r \tanh(\alpha t) + c, \ \frac{r}{\cosh(\alpha t)} \right)$$

with center the point $(c, o)$ and radius $r$ defined by

$$c = \frac{x_2^2 - x_1^2 + y_2^2 - y_1^2}{2(x_2 - x_1)}$$

and

$$r = \sqrt{(x_1 - c)^2 + y_1^2} = \sqrt{(x_2 - c)^2 + y_2^2}.$$

First we will find the value $t_0$ such that $\psi(t_0) = z_1$ and we will define a new curve $\phi$ by

$$\phi(t) := \psi(t + t_0),$$

so $\phi(0) = z_1$.

To find $t_0$ we have to solve the following system of equations

$$\begin{cases} r \tanh(\alpha t_0) + c = x_1 \\ \dfrac{r}{\cosh(\alpha t_0)} = y_1 \end{cases}.$$

This gives

$$t_0 = \frac{1}{\alpha} \operatorname{arccos} h\left( \frac{r}{y_1} \right) = \frac{1}{\alpha} \operatorname{arcsin} h\left( \frac{x_1 - c}{y_1} \right)$$

for every $\alpha \neq 0$. That is

$$\phi(t) = \psi\left(t + \frac{1}{\alpha}\operatorname{arccos}h\left(\frac{r}{y_1}\right)\right).$$

By imposing that

$$\phi(\tau) = \psi\left(\tau + \frac{1}{\alpha}\operatorname{arccos}h\left(\frac{r}{y_1}\right)\right) = z_2,$$

since

$$\psi\left(\frac{1}{\alpha}\operatorname{arccos}h\left(\frac{r}{y_2}\right)\right) = z_2$$

this implies that

$$\tau + \frac{1}{\alpha}\operatorname{arccos}h\left(\frac{r}{y_1}\right) = \frac{1}{\alpha}\operatorname{arccos}h\left(\frac{r}{y_2}\right)$$

and gives

$$\alpha = \frac{1}{\tau}\log\left(\frac{y_1}{y_2}\frac{r + \sqrt{r^2 - y_2^2}}{r + \sqrt{r^2 - y_1^2}}\right)$$

by using the relation

$$\operatorname{arccos}h(x) = \log\left(x + \sqrt{x^2 - 1}\right)$$

for $x \geq 1$ (this is possible since $r \geq y_1, y_2$).

Now suppose that $x_1 = x_2$. The geodesic containing $z_1 = (x, y_1)$ and $z_2 = (x, y_2)$ is a vertical line of the form

$$\psi(t) = \left(x, e^{\alpha t}\right).$$

If $t_0 = \frac{1}{\alpha}\log y_1$ then $\psi(t_0) = \left(x, e^{\alpha t}\right)$, so we set

$$\phi(t) := \psi(t + t_0) = \left(x, e^{\alpha(t+t_0)}\right).$$

We determine $\alpha$ in order to have $\phi(\tau) = z_2$; this gives

$$\alpha\tau + \log y_1 = \log y_2,$$

implying that

$$\alpha = \frac{1}{\tau} \log \frac{y_2}{y_1}.$$

# 4  The SABR model

## 4.1 RIEMANNIAN METRIC FOR THE SABR

Let $b, C$ and $\sigma$ be functions from $I\!R$ to $I\!R$ with $C$ integrable, let $F_0$ and $\alpha$ be real constants and let $P = (\rho_{ij})_{i,j=1,2}$ be a symmetric and invertible matrix of order $2$ whose diagonal entries are equal to $1$. A stochastic volatility model is defined as the following system of two correlated stochastic differential equations:

$$\begin{cases} F_t(\omega) = F_0 + \int_0^t A_u(\omega) C(F_u(\omega)) dW_u(\omega) \\ A_t(\omega) = \alpha + \int_0^t b(A_u(\omega)) du + \int_0^t \sigma(A_u(\omega)) dZ_u(\omega) \\ \qquad corr(W_t^i, Z_t^j) = \rho_{ij} \end{cases}$$

Set $\rho := \rho_{12}$; the inverse of the matrix $P$ is the matrix

$$\frac{1}{1-\rho^2}\begin{pmatrix} 1 & -\rho \\ -\rho & 1 \end{pmatrix},$$

in particular this implies that $\rho \neq 1$. For $i, j \in \{1,2\}$ we denote $\rho^{ij}$ the $(i, j)$-th element of $P^{-1}$. Set $\sigma_1(f, a) := a C(f)$, $\sigma_2(f, a) := \sigma(a)$, $z_1 := f$, $z_2 := a$ and let $p = p(t; f, a; f', a')$ be solution of the equation

$$\frac{\partial}{\partial t} p(t; f, a; f', a') = b(f, a) \frac{\partial p(t; f, a; f', a')}{\partial a} + \frac{1}{2}\left(\sum_{i,j=1}^2 \rho_{ij} \sigma_i(x) \sigma_j(x) \frac{\partial^2 p(t; f, a; f', a')}{\partial z_i \partial z_j}\right).$$

For $i, j \in \{1,2\}$ set

$$g^{ij}(f, a) := \frac{1}{2}\rho_{ij}\, \sigma_i(f, a)\, \sigma_j(f, a)$$

and suppose that the matrix $(g^{ij}(f,a))_{i,j=1,\cdots,n}$ is positive definite for every $(f,a) \neq (0,0)$. The Riemannian metric $g^{-1}$ for the above system of stochastic differential equations is

$$g^{-1}(f,a) := \frac{1}{2} \begin{pmatrix} a^2 C^2(f) & \rho a \sigma(a) C(f) \\ \rho a \sigma(a) C(f) & \sigma^2(a) \end{pmatrix}.$$

Let us consider the function

$$IR^2 \to IR^2$$

$$(f,a) \to \left( q(f) := \int_{f_0}^{f} \frac{df'}{C(f')}, \; \xi(a) := \int_{0}^{a} \frac{a'}{\sigma(a')} da' \right);$$

its Jacobian is

$$\begin{pmatrix} \frac{1}{C(f)} & 0 \\ 0 & \frac{a}{\sigma(a)} \end{pmatrix},$$

so $g^{-1}$ becomes

$$\frac{1}{2} \begin{pmatrix} \frac{1}{C(f)} & 0 \\ 0 & \frac{a}{\sigma(a)} \end{pmatrix} \begin{pmatrix} a^2 C^2(f) & \rho a \sigma(a) C(f) \\ \rho a \sigma(a) C(f) & \sigma^2(a) \end{pmatrix} \begin{pmatrix} \frac{1}{C(f)} & 0 \\ 0 & \frac{a}{\sigma(a)} \end{pmatrix} = \frac{a^2}{2} \begin{pmatrix} 1 & \rho \\ \rho & 1 \end{pmatrix}.$$

Moreover we have that

$$b^1(q,\xi) = 0, \qquad b^2(q,\xi) = \frac{a}{\sigma(a)} b(a).$$

Now consider the following change of coordinates,

$$F' : IR^2 \to IR^2$$

$$(q,\xi) \to \left( x(q,\xi) := q - \rho \xi, \; y(\xi) := \sqrt{1-\rho^2}\, \xi \right)$$

Its Jacobian is

$$\begin{pmatrix} 1 & -\rho \\ 0 & \sqrt{1-\rho^2} \end{pmatrix},$$

so $g^{-1}$ becomes

$$\frac{a^2}{2}\begin{pmatrix}1 & -\rho \\ 0 & \sqrt{1-\rho^2}\end{pmatrix}\begin{pmatrix}1 & \rho \\ \rho & 1\end{pmatrix}\begin{pmatrix}1 & 0 \\ -\rho & \sqrt{1-\rho^2}\end{pmatrix} = \frac{1-\rho^2}{2}a^2\begin{pmatrix}1 & 0 \\ 0 & 1\end{pmatrix}$$

and

$$b^1(x,y) = -\rho\frac{a}{\sigma(a)}b(a), \qquad b^2(x,y) = \sqrt{1-\rho^2}\,\frac{a}{\sigma(a)}b(a).$$

Now let us consider the SABR model, i.e., the stochastic volatility model with $b=0$ and $\sigma(z) = \nu z$ for some constant $\nu$. In particular this implies that $\xi(a) = \dfrac{a}{\nu}$ and $y(\xi(a)) = \dfrac{\sqrt{1-\rho^2}\,a}{\nu}$, so in particular

$$a(y) = \frac{\nu}{\sqrt{1-\rho^2}}\,y$$

and thus $g^{-1}$ in $(x, y)$–coordinates is

$$\frac{\nu^2}{2}y^2\begin{pmatrix}1 & 0 \\ 0 & 1\end{pmatrix},$$

so the second order derivatives term is

$$\frac{\nu^2}{2}y^2\left(\frac{\partial^2 p}{\partial x_1^2}(t, z_1, z_2) + \frac{\partial^2 p}{\partial x_2^2}(t, z_1, z_2)\right).$$

Now let us consider the function

$$\tau(t) := \frac{\nu^2}{2}t;$$

then

$$\frac{\partial p}{\partial t} = \frac{\partial p}{\partial \tau}\frac{\partial \tau}{\partial t} = \frac{\nu^2}{2}\frac{\partial p}{\partial \tau}$$

so

$$\frac{v^2}{2}\frac{\partial p}{\partial \tau}(\tau, z_1, z_2) = \frac{v^2}{2} y_1^2 \left( \frac{\partial^2 p}{\partial x_1^2}(\tau, z_1, z_2) + \frac{\partial^2 p}{\partial y_1^2}(\tau, z_1, z_2) \right)$$

and thus

$$\frac{\partial p}{\partial \tau}(\tau, z_1, z_2) = y_1^2 \left( \frac{\partial^2 p}{\partial x_1^2}(\tau, z_1, z_2) + \frac{\partial^2 p}{\partial y_1^2}(\tau, z_1, z_2) \right).$$

Thus

$$g^{-1}(z_1) = y_1^2 \begin{pmatrix} 1 & 0 \\ 0 & 1 \end{pmatrix}$$

and

$$g(z_1) = \frac{1}{y_1^2} \begin{pmatrix} 1 & 0 \\ 0 & 1 \end{pmatrix},$$

that is the metric is that of the Poincaré half-plane.

## 4.2 HEAT KERNEL EXPANSION FOR THE SABR

Let us set

$$F = F(x_1, y_1; x_2, y_2) := 1 + \frac{(x_2 - x_1)^2 + (y_2 - y_1)^2}{2 y_1 y_2};$$

then we have

$$\text{dist}\,(x_1, y_1; x_2, y_2) = \text{arccos}\,h\left( 1 + \frac{(x_2 - x_1)^2 + (y_2 - y_1)^2}{2 y_1 y_2} \right) = \text{arccos}\,h(F)$$

and thus

$$Syn\,(x_1, y_1; x_2, y_2) = \frac{1}{2} \text{arccos}\,h^2(F).$$

We have that

$$Van\,(x_1, y_1; x_2, y_2) = \frac{dist\,(x_1, y_1; x_2, y_2)}{\sinh(dist\,(x_1, y_1; x_2, y_2))}$$

(see Sec. 4.3 of [Paulot] or [5.92] of [Avramidi]); since for $z \geq 0$ we have

$$\sinh z = \sqrt{\cosh^2 z - 1},$$

then

$$\sinh(dist\,(x_1, y_1; x_2, y_2)) = \sinh(\arccos h(F)) =$$
$$= \sqrt{\cosh^2 \arccos h(F) - 1} = \sqrt{F^2 - 1},$$

then

$$Van\,(x_1, y_1; x_2, y_2) = \frac{\arccos h(F)}{\sqrt{F^2 - 1}}.$$

We have that

$$g(x, y) = \frac{1}{y^2}\begin{pmatrix} 1 & 0 \\ 0 & 1 \end{pmatrix}, \quad g^{-1}(x, y) = y^2 \begin{pmatrix} 1 & 0 \\ 0 & 1 \end{pmatrix},$$

then

$$g_{ih}(x, y) = \frac{\delta_{ih}}{y^2}, \quad g^{hk}(x, y) = \delta_{hk}\, y^2, \quad \sqrt{\det g(x, y)} = \frac{1}{y}.$$

Since for $i = 1, 2$ we have that

$$A_i(x, y) = \frac{1}{2}\sum_{h=1}^{2} g_{ih}(x, y)\left[ b^h(x, y) - \frac{\sum_{k=1}^{2} \partial_k\left(\sqrt{\det g(x, y)}\, g^{hk}(x, y)\right)}{\sqrt{\det g(x, y)}} \right]$$

and that $b^h = 0$, then

$$A_i(x, y) = \frac{1}{2}\sum_{h=1}^{2} \frac{\delta_{ih}}{y^2}\left[ -y \sum_{k=1}^{2} \partial_k\left( \frac{1}{x_2}\delta_{hk}\, y^2 \right) \right] = -\frac{1}{2y}\sum_{h,k=1}^{2} \delta_{ih}\, \partial_k(\delta_{hk}\, y) = -\frac{1}{2y}\partial_i y$$

and thus

$$A_1(x, y) = 0, \quad A_2(x, y) = -\frac{1}{2y}.$$

Now let us given two points $z_1 = (x_1, y_1)$ and $z_2 = (x_2, y_2)$ with $x_1 \neq x_2$. Let $t > 0$; as seen before, the geodesic

$$\phi : [0, t] \to IH^2$$

such that $\phi(0) = z_1$ and $\phi(t) = z_2$ is given by

$$\phi(u) = \left( r \tanh(\alpha u + u_0) + c, \ \frac{r}{\cosh(\alpha u + u_0)} \right)$$

with

$$c = \frac{x_2^2 - x_1^2 + y_2^2 - y_1^2}{2(x_2 - x_1)}, \quad r = \sqrt{(x_1 - c)^2 + y_1^2} = \sqrt{(x_2 - c)^2 + y_2^2},$$

$$u_0 = \operatorname{arccosh}\left(\frac{r}{y_1}\right) = \operatorname{arcsinh}\left(\frac{x_1 - c}{y_1}\right), \quad \alpha = \frac{1}{t} \log\left( \frac{y_1}{y_2} \frac{r + \sqrt{r^2 - y_2^2}}{r + \sqrt{r^2 - y_1^2}} \right).$$

We have that

$$Par_\phi(t) = \exp\left( -\sum_{i=1}^{2} \int_0^t \dot\phi_i(u) A_i(\phi(u)) du \right) = \exp\left( -\int_0^t \dot\phi_2(u) A_2(\phi(u)) du \right),$$

so we need to compute $\dot\phi_2(u)$. We have that

$$\dot\phi_2(u) = \frac{d}{du} \frac{r}{\cosh(\alpha u + u_0)} = \frac{-r\alpha \sinh(\alpha u + u_0)}{\cosh^2(\alpha u + u_0)},$$

then

$$Par_\phi(t) = \exp\left( -\int_0^t \dot\phi_2(u) A_2(\phi(u)) du \right) = \exp\left( -\frac{1}{2} \int_0^t \frac{r\alpha \sinh(\alpha u + u_0)}{\cosh^2(\alpha u + u_0)} \frac{\cosh(\alpha u + u_0)}{r} du \right) =$$

$$= \exp\left( -\frac{\alpha}{2} \int_0^t \frac{\sinh(\alpha u + u_0)}{\cosh(\alpha u + u_0)} du \right) = \exp\left( -\frac{\alpha}{2} \int_0^t \tanh(\alpha u + u_0) du \right).$$

By making the change of variables

$$v = \alpha u + u_0,$$

we have that

$$Par_\phi(t) = \exp\left(-\frac{1}{2}\int_0^{\alpha t+u_0} \tanh v\, dv\right) = \exp\left(-\frac{1}{2}[\log\cosh v]_0^{\alpha t+u_0}\right) =$$

$$= \exp\left(-\frac{1}{2}\log\cosh(\alpha t + u_0)\right) = \frac{1}{\sqrt{\cosh(\alpha t + u_0)}}.$$

Now let us given two points $z_1 = (x, y_1)$ and $z_2 = (x, y_2)$ and let $t > 0$; the geodesic

$$\phi:[0,t] \to IH^2$$

such that $\phi(0) = z_1$ and $\phi(t) = z_2$ is given by

$$\phi(u) = \left(x, e^{\alpha u + u_0}\right),$$

with $u_0 = \log y_1$ and $\alpha = \frac{1}{t}\log\frac{y_2}{y_1}$.

As before we have that

$$\dot\phi_2(u) = \alpha e^{\alpha u + u_0},$$

then

$$Par_\phi(t) = \exp\left(-\int_0^t \dot\phi_2(u) A_2(\phi(u))du\right) = \exp\left(\frac{\alpha}{2}\int_0^t e^{\alpha u + u_0}\frac{1}{e^{\alpha u + u_0}}du\right) = \exp\left(\frac{\alpha t}{2}\right).$$

We have that

$$Q(x, y) := \sum_{i,j=1}^{2}\left[g^{ij}(x,y)A_i(x,y)A_j(x,y) + \frac{\partial_i\left(\sqrt{\det g(x,y)}\, g^{ij}(x,y)A_j(x,y)\right)}{\sqrt{\det g(x,y)}}\right] =$$

$$= \sum_{i,j=1}^{2}\left[\delta_{ij}y^2 A_i(x,y)A_j(x,y) + y\partial_i\left(\frac{1}{y}\delta_{ij}y^2 A_j(x,y)\right)\right] =$$

$$= \sum_{i=1}^{2}\left[y^2 A_i^2(x,y) + y\partial_i(yA_i(x,y))\right] = y^2 A_2^2(x,y) + y\partial_y(yA_2(x,y)) =$$

$$= y^2 \frac{1}{4y^2} + y\partial_y\left(y\left(-\frac{1}{2y}\right)\right) = \frac{1}{4}.$$

We have that

$$\exp\left(-\frac{Syn(x_1, y_1; x_2, y_2)}{2t}\right) = \exp\left(-\frac{1}{4t}\operatorname{arccos} h^2(F)\right),$$

then for $x_1 \neq x_2$ we have

$$p(t; x_1, y_1; x_2, y_2) = \frac{1}{4\pi t y_2}\sqrt{\frac{\operatorname{arccos} h(F)}{\cosh(\alpha t + u_0)\sqrt{F^2-1}}} \exp\left(-\frac{1}{4t}\operatorname{arccos} h^2(F)\right) \sum_{k=0}^{+\infty} a_k(x_1, y_1; x_2, y_2) t^k$$

and we have also

$$p(t; x, y_1; x, y_2) = \frac{1}{4\pi t y_2}\sqrt{\frac{\operatorname{arccos} h(F_0)}{\cosh(\alpha t + u_0)\sqrt{F_0^2-1}}} \exp\left(\frac{\alpha t}{2} - \frac{1}{4t}\operatorname{arccos} h^2(F_0)\right) \sum_{k=0}^{+\infty} a_k(x, y_1; x, y_2) t^k$$

where

$$F_0 = 1 + \frac{(y_2 - y_1)^2}{2 y_1 y_2}.$$

We have that $a_0(x_1, y_1; x_2, y_2) = 1$ and that

$$a_1(x, y; x, y) = \frac{1}{6}R(x, y) + Q(x, y) = -\frac{1}{3y} + \frac{1}{4},$$

where

$$R(x, y) = -\frac{2}{y}$$

is the scalar curvature.

By using the approximation at first order of the preceding formula, one can find the following value for the BSM implied volatility

$$\sigma_{BS}(K,T) = \frac{\log\frac{K}{f_0}}{\int_{f_0}^{K} \frac{1}{\sqrt{2g^{ff}(a_{\min})}} dh} \times$$

$$\left(1 + \frac{g^{ff}(a_{\min})T}{12}\left(-\frac{3}{4}\left(\frac{\partial_f g^{ff}(a_{\min})}{g^{ff}(a_{\min})}\right)^2 + \frac{\partial_f^2 g^{ff}(a_{\min})}{g^{ff}(a_{\min})} + \frac{1}{f_{av}^2}\right) + \frac{g^{ff'}(a_{\min})T}{2g^{ff}(a_{\min})\phi''(a_{\min})}\left(\log(\Delta g P^2)'(a_{\min}) - \frac{\phi'''(a_{\min})}{\phi'''(a_{\min})} + \frac{g^{ff''}(a_{\min})}{g^{ff'}(a_{\min})}\right)\right)$$

The details of the derivation of such a formula are in 4.3.3 of [H-L 2005] or 6.1.4 of [H-L 2008].

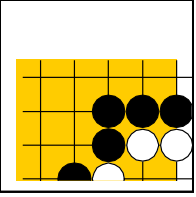